\documentclass[aps,twocolumn,showpacs,preprintnumbers,floats]{revtex4}\def\@cite#1#2{\textsuperscript{[{#1\if@tempswa , #2\fi}]}}
\usepackage{mathrsfs}
\usepackage{longtable,lscape}
\usepackage{txfonts}
\usepackage{amssymb}
\usepackage{indentfirst}
\usepackage{graphicx,booktabs}
\usepackage{color}
\usepackage{amssymb}
\usepackage{epsfig}
\newcommand{\vsig}{\mbox{\boldmath$\sigma$\unboldmath}}

\begin{document}

\title{Strong decays of higher excited heavy-light mesons in a chiral quark model}
\author{
Li-Ye Xiao and Xian-Hui Zhong \footnote {zhongxh@hunnu.edu.cn}}
\affiliation{ Department of Physics, Hunan Normal University, and
Key Laboratory of Low-Dimensional Quantum Structures and Quantum
Control of Ministry of Education, Changsha 410081, China }


\begin{abstract}

The strong decay properties of the higher excited heavy-light mesons
from the first radially excited states up to the first $F$-wave
states are studied in a constituent quark model. It is found that
many missing excitations have good potentials to be found in future
experiments for their narrow widths, and some of them dominantly
decay into the first orbital excitations rather than into ground
states. In future observations, one should focus on the decay
processes not only into the ground states, but also into the
low-lying $P$-wave excitations with $J^P=0^+,1^+$. Furthermore, the
nature of the newly observed states $D_J(3000)$, $D_J^*(3000)$ and
$B(5970)$ is discussed. It is predicted that $D_J(3000)$ seems to be
a partner of $D_{sJ}(3040)$, which could be identified as the
low-mass mixed state $|2{P_1}\rangle_L$ ($J^P=1^+$) via the
$2^1P_1$-$2^3P_1$ mixing. The $D_J^*(3000)$ resonance seems to favor
the $1^3F_4$ state, while the quantum numbers $J^P=0^+$ and $2^+$
cannot be excluded complectly, more experimental observations are
needed to determine its $J^P$ values. The $B(5970)$ resonance is
most likely to be the $1^3D_3$ with $J^P=3^-$.

\end{abstract}


\maketitle

\section{Introduction}

During the past several years, significant progress has been made in
the observation of the heavy-light mesons. More and more higher
excitations have been found in experiments. In the $D$-meson family,
four new members, the $D(2550)$, $D(2600)$, $D(2750)$ and $D(2760)$
were discovered by the $BABAR$
Collaboration~\cite{delAmoSanchez:2010vq}, which were confirmed by
the LHCb Collaboration with slightly different
masses~\cite{Aaij:2013sza}. Furthermore, the LHCb Collaboration
observed two new higher $D$-meson excitations, $D^*_J(3000)$ and
$D_J(3000)$, with natural and unnatural parities,
respectively~\cite{Aaij:2013sza}. In the $D_s$-meson family, except
for the $D_{s1}(2700)$ and $D_{sJ}(2860)$ observed by $BABAR$ and
Belle before~\cite{Aubert:2006mh,jb:2007aa}, recently a new broad
higher $D_s$-meson excitation $D_{sJ}(3040)$ was observed by $BABAR$
as well~\cite{Aubert:2009ah}. Very recently, progress has been
obtained in the search for the excited $B$ meson as well. The CDF
Collaboration observed some evidence of a new narrow excited state
$B(5970)$ in the $B\pi$ channel~\cite{Aaltonen:2013atp}.

About the newly observed states $D(2550)$, $D(2600)$, $D(2750)$,
$D(2760)$, $D_{s1}(2700)$, $D_{sJ}(2860)$ and $D_{sJ}(3040)$, their
strong decays had been analyzed in our previous
papers~\cite{Zhong:2008kd,Zhong:2010vq,Zhong:2009sk}. We found that
the $D(2760)$ could be identified as the $1^3D_3$ with $J^P=3^-$,
while the $D(2750)$ is most likely to be the mixed state
$|1{D_2}'\rangle_H$ ($J^P=2^-$) via the $1^1D_2$-$1^3D_2$ mixing.
The $D(2600)$ favors the mixed state $|(SD)_1\rangle_L$ $(J^P=1^-)$
via the $1^3D_1$-$2^3S_1$ mixing. The $D(2550)$ might be assigned as
the $2^1S_0$ assignment although its predicted width is obviously
narrower than that given in experiments. For the newly observed
$D_{sJ}$ mesons, we found that the $D_{s1}(2700)$ is most likely to
be the strange partner of the $D(2600)$. There might exist two-state
scenario for the $D_{sJ}(2860)$, one resonance corresponds to the
strange partner of the $D(2760)$ [denoted by $D_{sJ_1}(2860)$] and
the other resonance is the strange partner of the $D(2750)$ [denoted
by $D_{sJ_2}(2860)$]. The $D_{sJ}(3040)$ could be identified as the
low-mass physical state $|2P_1\rangle_L$ via the $2^1P_1$-$2^3P_1$
mixing. More discussions of these newly observed $D$ and $D_s$
mesons can be found in the
literature~\cite{Lu:2014zua,Godfrey:2013aaa,Zhen:2013dca,Wang:2012wk,Colangelo:2007ds,Colangelo:2012xi,Mohler:2012na,Badalian:2011tb,Mohler:2011ke,Chen:2011rr,Wang:2010ydc,vanBeveren:2009jq,Sun:2009tg,Guo:2011dd,Sun:2010pg,
Colangelo:2010te,Li:2009qu,Close:2005se,Close:2006gr,Swanson:2006st,Zhang:2006yj}.

On the other hand, about the higher excitations $D^{*}_J(3000)$,
$D_J(3000)$ and $B(5970)$ observed very recently, a few studies can
be found in the
literature~\cite{Xu:2014mqa,Yu:2014dda,Sun:2014wea,Sun:2013qca,Wang:2013tka,Xu:2014mka,Wang:2014cta}.
For example, the $D^{*}_J(3000)$ and $D_J(3000)$ states were
explained as the first $P$-wave states in the $D$ meson family with
$J^P=0^+$ and $1^+$, respectively~\cite{Sun:2010pg}. As a candidate
of the $2^3S_1$~\cite{Xu:2014mka,Wang:2014cta}, $1^3D_1$ and
$1^3D_3$ states~\cite{Wang:2014cta}, the strong decay properties of
the $B(5970)$ were studied with the heavy meson effective theory.

These newly observed higher excitations provide us a good chance to
establish a fairly abundant $D$- and $D_s$-meson spectroscopy, which
has been summarized in Tab.~\ref{SPE}. For comparisons, some
interested predictions of the heavy-light meson spectroscopy in
theory are also included~\cite{Di Pierro:2001uu,Ebert:2009ua}. From
the table, it is seen that the knowledge about the $B$- and
$B_s$-meson spectroscopy is very poor. Except for the ground states,
only two low-lying orbital excitations in the $P$ waves are found in
the $B$- and $B_s$-meson spectroscopy. The discovery of the
$B(5970)$ enhances our confidence in the search for more higher
excitations in the $B$- and $B_s$-meson families. From now on, we
might have a golden time to study the higher excitations of the
heavy-light mesons.

To provide helpful information for the experimental search for more
excited heavy-light mesons and to gain a unified understanding of
the newly observed resonances, in this work we continue to study the
strong decay properties of the higher excitations of heavy-light
mesons up to the $F$ waves (see Tab.~\ref{SPE}) with the chiral
quark model as well. This model has been developed and successfully
used to deal with the strong decays of heavy-light mesons, charmed
and strange baryons~\cite{Zhong:2008kd,Zhong:2010vq,Zhong:2009sk,
Zhong:2007gp,Liu:2012sj,Xiao:2013xi}.

This paper is organized as follows. In the subsequent section, a
brief review of the model is given. The numerical results are
presented and discussed in Sec.~\ref{results}. Finally, a summary is
given in Sec.\ \ref{suma}.

\begin{widetext}
\begin{center}
\begin{table}[ht]
\caption{The heavy-light meson spectroscopy predicted in theory
compared with the observations in experiments. The predicted masses
(MeV) are obtained from~\cite{Di Pierro:2001uu,Ebert:2009ua}, while
the observed states are obtained from the
PDG~\cite{Beringer:1900zz}. The $1P_1$ and $1P_1'$ stand for the
low-mass and high-mass mixed states via the $1^1P_1$-$1^3P_1$
mixing, respectively, which have been defined
in~\cite{Zhong:2008kd}. The $1D_2$ and $1D_2'$ stand for the
low-mass and high-mass mixed states via the $1^1D_2$-$1^3D_2$
mixing, respectively. The $2P_1$ and $2P_1'$ stand for the low-mass
and high-mass mixed states via the $2^1P_1$-$2^3P_1$ mixing,
respectively. }\label{SPE}
\begin{tabular}{c|cc|cc|cc|cc|cc}  \hline\hline
 States&\multicolumn{2}{|c|}{\underline{ $D$ mesons }}& \multicolumn{2}{|c|}{\underline{ $D_s$ mesons }} &\multicolumn{2}{|c|}{\underline{$B$ mesons }} &\multicolumn{2}{|c|}{\underline{ $B_s$ mesons }}
 \\ 
 $n^{2S+1}L_J$&  Predicted Mass  & Observed state & Predicted Mass  &Observed state & Predicted Mass  &Observed state & Predicted Mass  & Observed state
\\ \hline
$1^1S_0$           &1868/1871&$D  (1865)$  \ \  &1965/1969\ \  &$D_s(1969)$         &5279/5280&$B(5280)$   \ \  &5373/5372 \ \  &$B_s(5367)$\\
$1^3S_1$           &2005/2010&$D^*(2007)$  \ \  &2113/2111 \ \ &$D^*_{s}(2112)$   &5324/5326&$B^*(5325)$ \ \  &5421/5414 \ \  &$B^*_s(5415)$\\
$1^3P_0$           &2377/2406&$D^*_0(2400)$ \ \ &2487/2509 \ \ &$D_{s0}(2317)$    &5706/5749 & ?         \ \  &5804/5833  \ \  &?\\
$1P_1$             &2417/2426&$D_1(2430)$  \ \  &2535/2536  \ \ &$D_{s1}(2460)$    &5700/5723 & $B_{J}^*(5732)$?\  &5805/5831 \ \  &?\\
$1P_1'$            &2490/2469&$D_1(2420)$  \ \  &2605/2574\ \ &$D_{s1}(2536)$    &5742/5774& $B_{1}(5721)$ \ \ &5842/5865 \ \  &$B_{s1}(5830)$\\
$1^3P_2$           &2460/2460& $D_2(2460)$\ \   &2581/2571 \ \ &$D_{s2}(2573)$    &5714/5741& $B_{2}(5747)$\ \  &5820/5842  \ \  &$B_{s2}(5840)$\\
$2^1S_0$           &2589/2581& $D(2550)$?\ \   &2700/2688 \ \  &?                 &5886/5890& ? \ \              &5985/5976 \ \  &?\\
$2^3S_1$           &2692/2632& $D(2600)$?\ \   &2806/2731 \ \  &$D_{s1}(2700)$?    &5920/5906& ? \ \              &6019/5992 \ \  &?\\
$1^3D_1$           &2795/2788& ?           \ \ &2913/2913 \ \  &?                 &6025/6119& ? \ \           &6127/6209 \ \  &?\\
$1D_2$             &2775/2806& ?          \ \   &2900/2931 \ \  &?                 &5985/6102& ? \ \           &6095/6189 \ \  & ?\\
$1D_2'$            &2883/2850& $D(2750)$?         \ \ &2953/2961\ \    &$D_{sJ_2}(2860)$?  &6037/6121& ? \ \           &6140/6218\ \  &?\\
$1^3D_3$           &2799/2863& $D(2760)$?\ \  &2925/2971 \ \   &$D_{sJ_1}(2860)$?  &5993/6091& $B(5970)$? \ \  &6103/6191 \ \  &?\\
$2^3P_0$           &2949/2919& ? \ \          &3067/3054\ \    &?                 &6163/6221& ?  \ \          &6264/6318\ \  &?\\
$2P_1$             &2995/2932& $D_J(3000)$?   &3114/3067\ \    &$D_{sJ}(3040)$?   &6175/6209& ? \ \            &6278/6321 \ \  &?\\
$2P_1'$            &3045/3021& ?\ \           &3165/3154 \ \   &  ?              &6194/6281& ? \ \            &6296/6345\ \  &?\\
$2^3P_2$           &3035/3012& ?\ \           &3157/3142\ \    &?                &6188/6260& ? \ \            &6292/6359 \ \  &?\\
$1^3F_2$           &3101/3090& ?\ \           &3224/3230\ \    &?                &6264/6412& ? \ \            &6369/6501 \ \   &?\\
$1F_3$             &3074/3129& ?\ \           &3203/3254\ \    &?                &6220/6391& ? \ \            &6332/6468 \ \  &?\\
$1F_3'$            &3123/3145& ?\ \           &3247/3266\ \    &?                &6271/6420& ? \ \            &6376/6515 \ \  &?\\
$1^3F_4$           &3091/3187& ?\ \           &3220/3300\ \    &?                &6226/6380& ? \ \            &6337/6475 \ \  &?\\
 \hline \hline
\end{tabular}
\end{table}
\end{center}
\end{widetext}

\section{The model}\label{model}

In this section, we give a brief review of the chiral quark model.
The details of this model can be found in our previous
papers~\cite{Zhong:2008kd,Zhong:2010vq,Zhong:2009sk}. In this model,
the low energy quark-pseudoscalar-meson and quark-vector-meson
interactions in the SU(3) flavor basis might be described by the
effective Lagrangians
\begin{eqnarray}
{\cal L}_{Pqq}&=&\sum_j
\frac{1}{f_m}\bar{\psi}_j\gamma^{j}_{\mu}\gamma^{j}_{5}\psi_j\partial^{\mu}\phi_m,\label{coup}\\
{\cal
L}_{Vqq}&=&\sum_j\bar{\psi}_j(a\gamma^{j}_{\mu}+\frac{ib}{2m_j}\sigma_{\mu\nu}q^\nu)V^\mu\psi_j
\label{coup2},
\end{eqnarray}
respectively, where $\psi_j$ represents the $j$th quark field in the
hadron, $\phi_m$ is the pseudoscalar meson field, $f_m$ is the
pseudoscalar meson decay constant, and $V^\mu$ represents the vector
meson field. Parameters $a$ and $b$ denote the vector and tensor
coupling strength, respectively.

In this model, the wave function of a heavy-light meson is adopted
by the nonrelativistic harmonic oscillator wave function, i.e.,
$\psi^n_{lm}=R_{nl}Y_{lm}$. To match the nonrelativistic wave
functions of the heavy-light mesons, we should adopt the
nonrelativistic form of Eq.~(\ref{coup}) in the calculations, which
is given by~\cite{Li:1994cy,Li:1997gda,qk3}
\begin{eqnarray}\label{ccpk}
H_{m}=\sum_j\left[A \vsig_j \cdot \textbf{q}
+\frac{\omega_m}{2\mu_q}\vsig_j\cdot \textbf{p}_j\right]I_j
\varphi_m,
\end{eqnarray}
in the center-of-mass system of the initial meson, where we have
defined $A\equiv -(1+\frac{\omega_m}{E_f+M_f})$. On the other hand,
from Eq.~(\ref{coup2}), one can easily derive the nonrelativistic
transition operators for the emission of a transversely or
longitudinally polarized vector meson, which are given by
~\cite{zhao:1998fn,Zhao:2000tb,Zhao:2001jw}
\begin{eqnarray}\label{vc}
H_m^T=\sum_j \left\{i\frac{b'}{2m_q}\vsig_j\cdot
(\mathbf{q}\times\mathbf{\epsilon})+\frac{a}{2\mu_q}\mathbf{p}_j\cdot
\mathbf{\epsilon}\right\}I_j\varphi_m,
\end{eqnarray}
and
\begin{eqnarray}
H_m^L=\sum_j \frac{a M_v}{|\mathbf{q}|}I_j\varphi_m \ .
\end{eqnarray}
In the above equations,  $\textbf{q}$ and $\omega_m$ are the
three-vector momentum and energy of the final-state light meson,
respectively; $\textbf{p}_j$ is the internal momentum operator of
the $j$th quark in the heavy-light meson rest frame; $\vsig_j$ is
the spin operator corresponding to the $j$th quark of the
heavy-light system; and $\mu_q$ is a reduced mass given by
$1/\mu_q=1/m_j+1/m'_j$ with $m_j$ and $m'_j$ for the masses of the
$j$th quark in the initial and final mesons, respectively. $M_v$ is
the mass of the emitted vector meson. The plane wave part of the
emitted light meson is $\varphi_m=e^{-i\textbf{q}\cdot
\textbf{r}_j}$, and $I_j$ is the flavor operator defined for the
transitions in the SU(3) flavor space
\cite{Li:1997gda,qk3,zhao:1998fn,Zhao:2000tb,Zhao:2001jw}. The
parameter $b'$ in Eq.~(\ref{vc}) is defined as $b'\equiv b-a$.

For a light pseudoscalar meson emission in heavy-light meson strong
decays, the partial decay width can be calculated
with~\cite{Zhong:2008kd,Zhong:2007gp}
\begin{equation}\label{dww}
\Gamma=\left(\frac{\delta}{f_m}\right)^2\frac{(E_f+M_f)|\textbf{q}|}{4\pi
M_i(2J_i+1)} \sum_{J_{iz},J_{fz}}|\mathcal{M}_{J_{iz},J_{fz}}|^2 ,
\end{equation}
where $\mathcal{M}_{J_{iz},J_{fz}}$ is the transition amplitude, and
$J_{iz}$ and $J_{fz}$ stand for the third components of the total
angular momenta of the initial and final heavy-light mesons,
respectively. $\delta$ as a global parameter accounts for the
strength of the quark-meson couplings. It has been determined in our
previous study of the strong decays of the charmed baryons and
heavy-light mesons~\cite{Zhong:2007gp,Zhong:2008kd}. Here, we take
the same value as that determined in
Refs.~\cite{Zhong:2008kd,Zhong:2007gp}, i.e., $\delta=0.557$.

In the calculation, the standard quark model parameters are adopted.
Namely, we set $m_u=m_d=330$ MeV, $m_s=450$ MeV, $m_c=1700$ MeV, and
$m_b=5100$ MeV for the constituent quark masses. The harmonic
oscillator parameter $\beta$ in the wave functions of heavy-light
mesons is taken as $\beta=0.40$ GeV. The decay constants for $\pi$,
$K$ and $\eta$ mesons are taken as $f_{\pi}=132$ MeV,
$f_K=f_{\eta}=160$ MeV, respectively. For the quark-vector-meson
coupling strength which still suffers relatively large
uncertainties, we adopt the values extracted from vector meson
photoproduction, i.e. $a\simeq -3$ and $b'\simeq
5$~\cite{zhao:1998fn,Zhao:2000tb,Zhao:2001jw}. The masses of the
mesons used in the calculations are adopted from the
PDG~\cite{Beringer:1900zz}. With these parameters, the strong decay
properties of the well-known heavy-light mesons and charmed baryons
have been described
reasonably~\cite{Zhong:2008kd,Zhong:2010vq,Zhong:2009sk,
Zhong:2007gp,Liu:2012sj}.

\section{Calculations and Results }\label{results}

\subsection{The $2^1S_0$ states}

Recently, the $BABAR$ Collaboration observed an excited $D$-meson
state $D(2550)$ in the $D^{*}\pi$ channel with a width of
$\Gamma\simeq 130$ MeV~\cite{delAmoSanchez:2010vq}, which was also
observed by the LHCb Collaboration recently~\cite{Aaij:2013sza}. In
theory, the first radially excited $D$-meson state $2^1S_0$ has a
mass $\sim 2.58$ GeV~\cite{Godfrey:1985xj,Di
Pierro:2001uu,Ebert:2009ua}. Furthermore, the decay mode and the
\emph{BABAR} analysis of angle distributions indicate that the
$D(2550)$ should be classified as the radially excited state
$2^1S_0$. We have studied the strong decays of the $D(2550)$ as the
first radially excited $D$-meson state $2^1S_0$ in
Refs.~\cite{Zhong:2010vq,Zhong:2008kd}. The strong decays of
$D(2^1S_0)$ are dominated by the $D^*\pi$ and $D_0(2400)\pi$ modes.
In the present work we have improved our predictions according to
the new observations from the LHCb
Collaboration~\cite{Aaij:2013sza}. The results are listed in
Tab.~\ref{Tab:D21S0}. The predicted partial decay width ratio is
\begin{eqnarray}
\frac{\Gamma[D^*\pi]}{\Gamma[D_0(2400)\pi]}\simeq 0.28,
\end{eqnarray}
which strongly depends on the mass of $D_0(2400)$. Our predicted
width, $\Gamma\simeq 68$ MeV, is about a factor of 2 narrower than
the data. The $^3P_0$ model~\cite{Sun:2010pg} and relativistic quark
model~\cite{Di Pierro:2001uu} calculations also obtained a narrow
width for the $D(2^1S_0)$ state. 

If the $D(2550)$ is the $2^1S_0$ assignment indeed, the same
excitation in the $D_s$-, $B$-, and $B_s$-meson spectroscopy might
be found in future experiments. To obtain more information about
these missing states, we further study the strong decay properties
of these radially excited states $D_s(2^1S_0)$, $B(2^1S_0)$ and
$B_s(2^1S_0)$ in the following.

\begin{table}[ht]
\caption{The strong decay width (MeV) of the $D(2550)$ as the
$2^1S_0$ with a mass of 2580 MeV.} \label{Tab:D21S0}
\begin{tabular}{c|cccccc }\hline
\hline
  Mode  &$D^*\pi$ &$D_0(2400)\pi$ & Total    \\
\hline
 Width &15 & 53    & 68                    \\
\hline
\end{tabular}
\end{table}


In the $D_s$-meson family, the predicted mass of the first radially
excited state $2^1S_0$ is around $2.7$ GeV~\cite{Godfrey:1985xj,Di
Pierro:2001uu,Ebert:2009ua}. There is no experimental information
about this state. The strong decay properties of $D_s(2^1S_0)$ have
been studied in our previous work~\cite{Zhong:2008kd}. Its strong
decays are governed by the $D^*K$ channel. Based on our model, the
$D_s(2^1S_0)$ is a very narrow state with a width of $\Gamma\simeq
10$ MeV. Such a narrow state should be observed in the $D^*K$ final
states if it indeed exists.


In the $B$-meson family, the predicted mass of $B(2^1S_0)$ is around
$5.9$ GeV~\cite{Godfrey:1985xj,Di Pierro:2001uu,Ebert:2009ua}. We
calculate its strong decay properties, which are listed in
Tab.~\ref{Tab:B21S0}. It is seen that the strong decays are
dominated by the $B^*\pi$ and $B(1^3P_0)\pi$ channels, and the
predicted partial width ratio is
\begin{eqnarray}
\frac{\Gamma[B^*\pi]}{\Gamma[B(1^3P_0)\pi]}\simeq 1.3.
\end{eqnarray}
The $B(2^1S_0)$ is also a narrow state with a width of $\Gamma\simeq
40$ MeV, which is slightly narrower than that of $D(2^1S_0)$. The
decay width predicted by us is roughly compatible with the recent
calculations with $^3P_0$ model~\cite{Sun:2014wea}.

\begin{table}[ht]
\caption{The strong decay width (MeV) of the $B(2^1S_0)$ with a mass
of 5886 MeV. The mass of $B(1^3P_0)$ is taken as 5706 MeV.}
\label{Tab:B21S0}
\begin{tabular}{c|cccccc }\hline
\hline
Mode   &$B^*\pi$ &$B(1^3P_0)\pi$ &$B(1^3P_2)\pi$ &$B^*\eta$ &Total\\
\hline
Width &16 & 24       &0.01              &0.25       &40\\
\hline
\end{tabular}
\end{table}


In the $B_s$-meson family, the predicted mass of $B_s(2^1S_0)$ is
around 6.0 GeV~\cite{Godfrey:1985xj,Di Pierro:2001uu,Ebert:2009ua}.
We calculate the strong decay properties of this state with a mass
of $M=5985$ MeV. This state dominantly decays into the $B^*K$
channel. The total decay width is $\Gamma\simeq 26$ MeV, which is a
little narrower than that of the $^3P_0$
calculations~\cite{Sun:2014wea}.

As a whole, the first radially excited states $2^1S_0$ in the $D$-,
$D_s$-, $B$- and $B_s$-meson spectroscopy might have a fairly narrow
width. However, up to now none of them has been established. It is
still a puzzle why such narrow radially excited states are still
missing. More theoretical and experimental studies are needed.

\begin{widetext}
\begin{center}
\begin{figure}[ht]
\centering \epsfxsize=7.0 cm \epsfbox{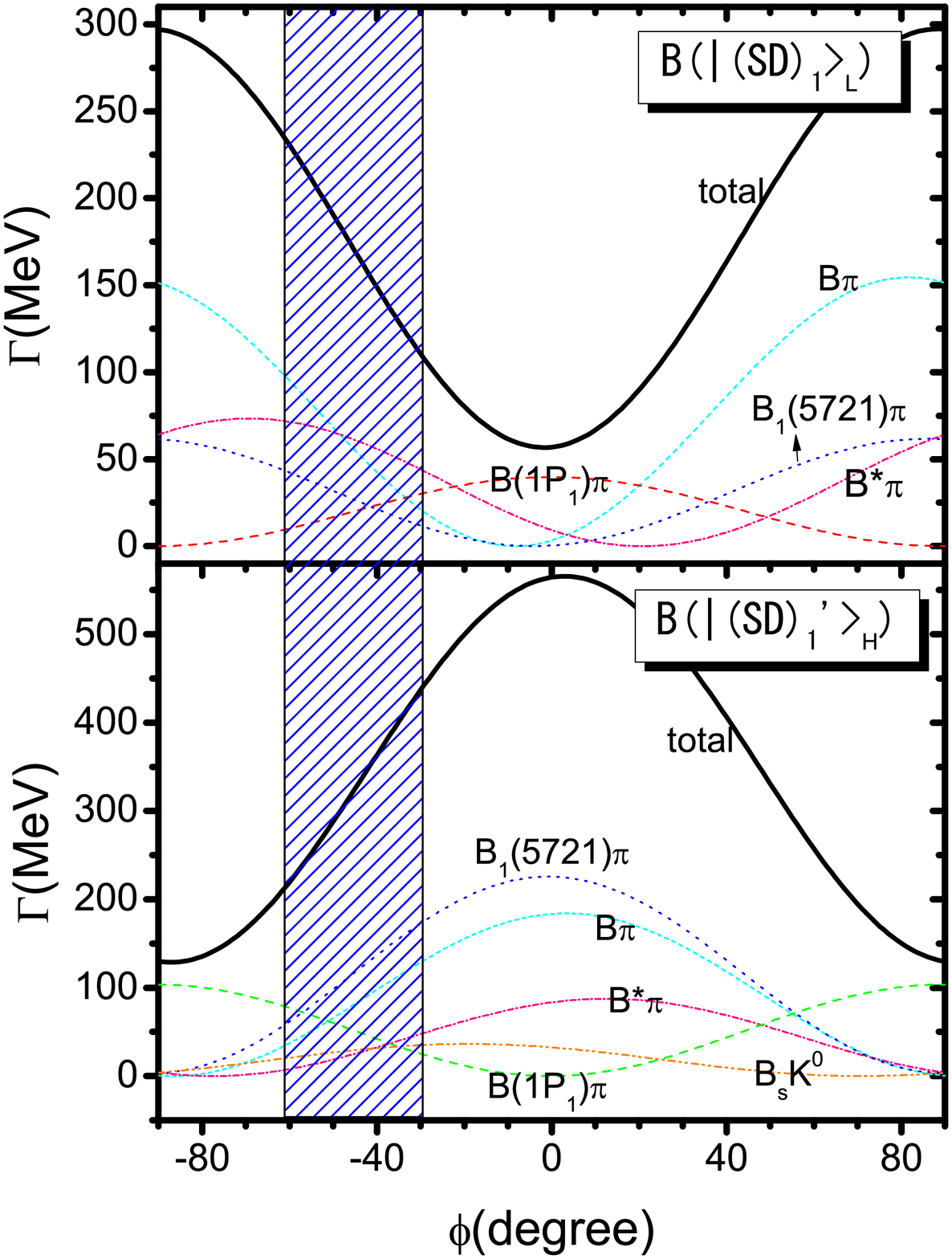} \epsfxsize=7.0 cm
\epsfbox{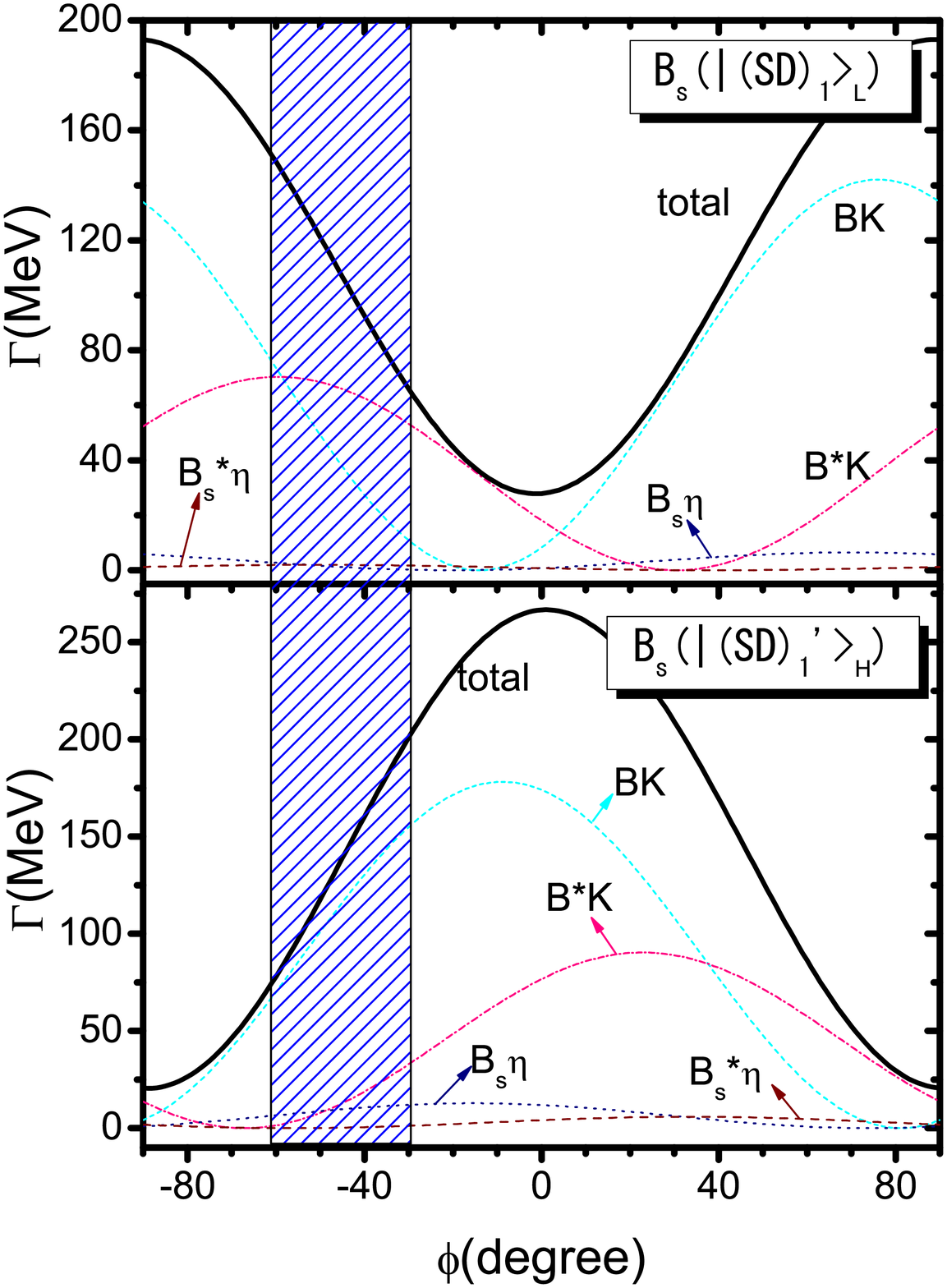} \caption{ The partial decay width and total decay
width for the mixed states via $2^3S_1$-$1^3D_1$ mixing in the $B$-
and $B_s$-meson families as functions of the mixing angle $\phi$.
The shaded bands correspond to the possible mixing angle region
derived from the strong decay properties of the $D(2600)$ and
$D_{s1}(2700)$. The masses for the mixed states in the $B$-meson
family, $B(|(SD)_{1}\rangle_L)$ and $B(|(SD)_{1}'\rangle_H)$, are
adopted as $5920$ and $6025$ MeV, respectively, while, the masses
for the mixed states in the $B_s$-meson family,
$B_s(|(SD)_{1}\rangle_L)$ and $B_s(|(SD)_{1}'\rangle_H)$, are
adopted as $6000$ and $6100$ MeV, respectively. The mass of
$B(1P_1)$ is taken as 5720 MeV. In the figure, we have hidden some
decay channels because of their small partial decay
widths.}\label{fig-SDB}
\end{figure}
\end{center}
\end{widetext}

\subsection{The $2^3S_1$-$1^3D_1$ mixing }

In 2010, the \emph{BABAR} Collaboration discovered a new excited
$D$-meson state $D(2600)$ in $D\pi$ and $D^*\pi$ decay channels with
a mass of $M=2612\pm6$ MeV and a total decay width of
$\Gamma=93\pm6\pm13$ MeV~\cite{delAmoSanchez:2010vq}, which might be
a partner of the $D_{s1}(2700)$ observed previously by \emph{BABAR}
and Belle~\cite{Aubert:2006mh,jb:2007aa}. The $D(2600)$ may
correspond to the $D(2650)$ observed by LHCb very
recently~\cite{Aaij:2013sza}.

In our previous work~\cite{Zhong:2010vq,Zhong:2009sk}, we have
carefully studied the strong decay properties of the $D(2600)$ and
$D_{s1}(2700)$. According to our analysis, both $D(2600)$ and
$D_{s1}(2700)$ could be explained as the mixed state
$|(SD)_1\rangle_L$ via the $2^3S_1$-$1^3D_1$ mixing:
\begin{equation}\label{mixsd}
\left(\begin{array}{c}|(SD)_1\rangle_L\cr |(SD)'_1\rangle_H
\end{array}\right)=\left(\begin{array}{cc} \cos\phi &\sin\phi\cr -\sin\phi & \cos\phi
\end{array}\right)
\left(\begin{array}{c} |2^3S_1\rangle \cr |1^3D_1\rangle
\end{array}\right).
\end{equation}
The mixing angle for the $D(2600)$ is $\phi\simeq-(36\pm 6)^\circ$,
while that for the $D_{s1}(2700)$ is $\phi=(-54\pm7)^\circ$. To
explain the strong decay properties of the $D(2600)$ and/or
$D_{s1}(2700)$, configuration mixing between $2^3S_1$ and $1^3D_1$
is also suggested in the
literature~\cite{Close:2006gr,Chen:2011rr,Li:2009qu}.

Following the mixing scheme in Eq.~(\ref{mixsd}), the strong decay
properties of the physical partners of the $D(2600)$ and
$D_{s1}(2700)$, i.e., $|(SD)_{1}'\rangle_H$, have been obtained as
well. Taking the mass of $D(|(SD)_{1}'\rangle_H)$ in the range of
$(2.65\sim 2.80)$ GeV, we predict that the $D(|(SD)_{1}'\rangle_H)$
is a broad state with a width of $\Gamma\simeq (360\pm 120)$ MeV.
Its strong decays are dominated by $D\pi$ and $D_1(2420)\pi$. On the
other hand, estimating the mass of $D_s(|(SD)_{1}'\rangle_H)$ in the
range of $(2.72\sim2.88)$ GeV, the total width is $\Gamma\simeq
(120\pm 10)$ MeV~\cite{Zhong:2009sk}. The predicted width for the
$D_s(|(SD)_{1}'\rangle_H)$ is not very broad. This state might be
observed in the $DK$ channel. It is interestingly found that the
LHCb Collaboration has observed a new resonance $D_{s1}(2860)$ with
a width of $\Gamma=(159\pm 122)$ MeV very
recently~\cite{Aaij:2014xza,Aaij:2014baa}. Their analysis shows that
the spin-parity should be $J^P=1^-$. This newly observed resonance
$D_{s1}(2860)$ can be favorably assigned as the physical partners of
the $D_{s1}(2700)$, i.e., $D_s(|(SD)_{1}'\rangle_H)$, predicted in
our previous work~\cite{Zhong:2009sk}.

If both $D(2600)$ and $D_{s1}(2700)$ indeed correspond to the mixed
state $|(SD)_1\rangle_L$, the $2^3S_1$-$1^3D_1$ mixing might be a
common character in the heavy-light mesons. The future search for
these missing mixing states $|(SD)_1\rangle_L$ and
$|(SD)_{1}'\rangle_H$ in the heavy-light meson spectroscopy becomes
interesting and important.

\begin{widetext}
\begin{center}
\begin{figure}[ht]
\centering \epsfxsize=7.0 cm \epsfbox{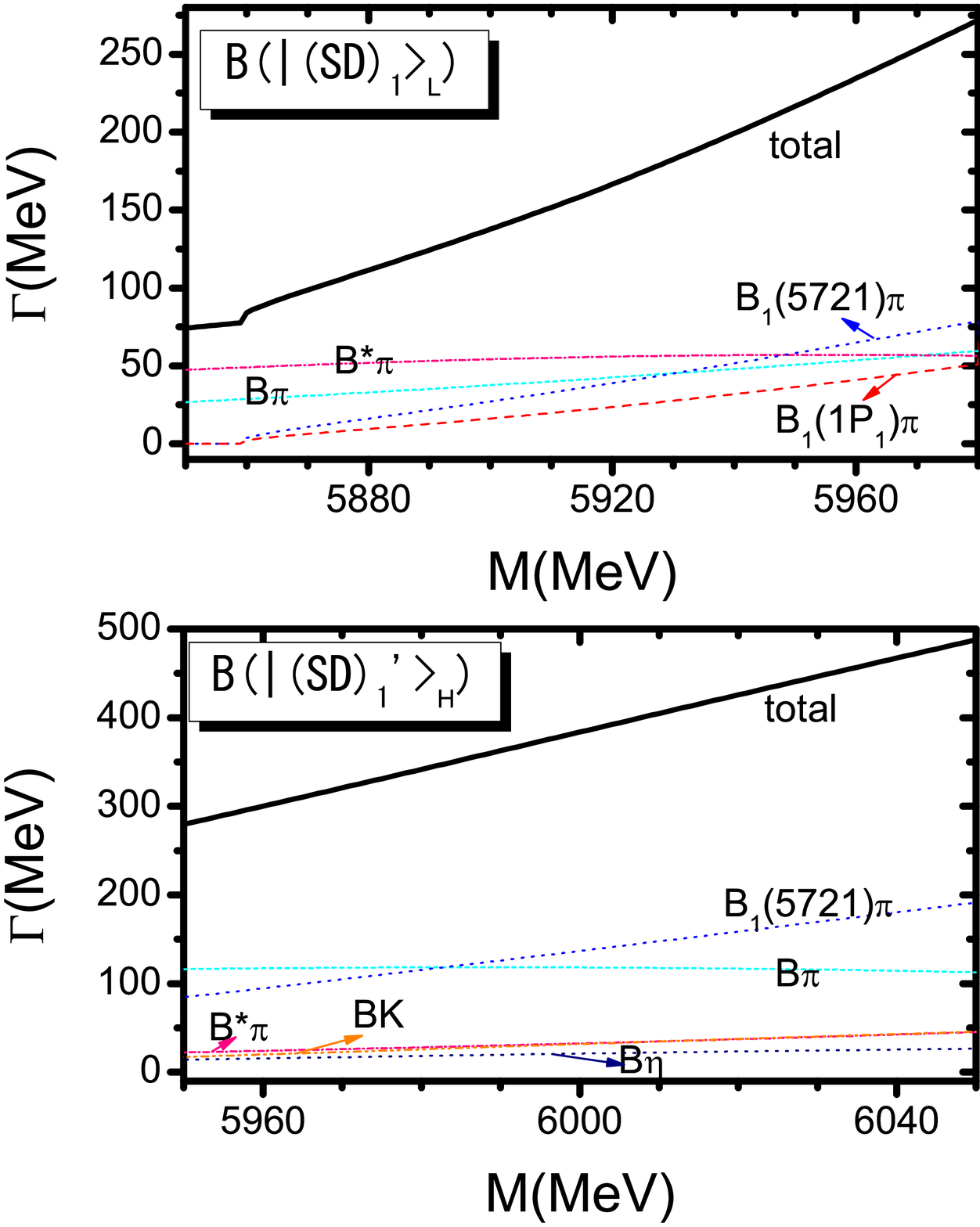} \epsfxsize=7.0 cm
\epsfbox{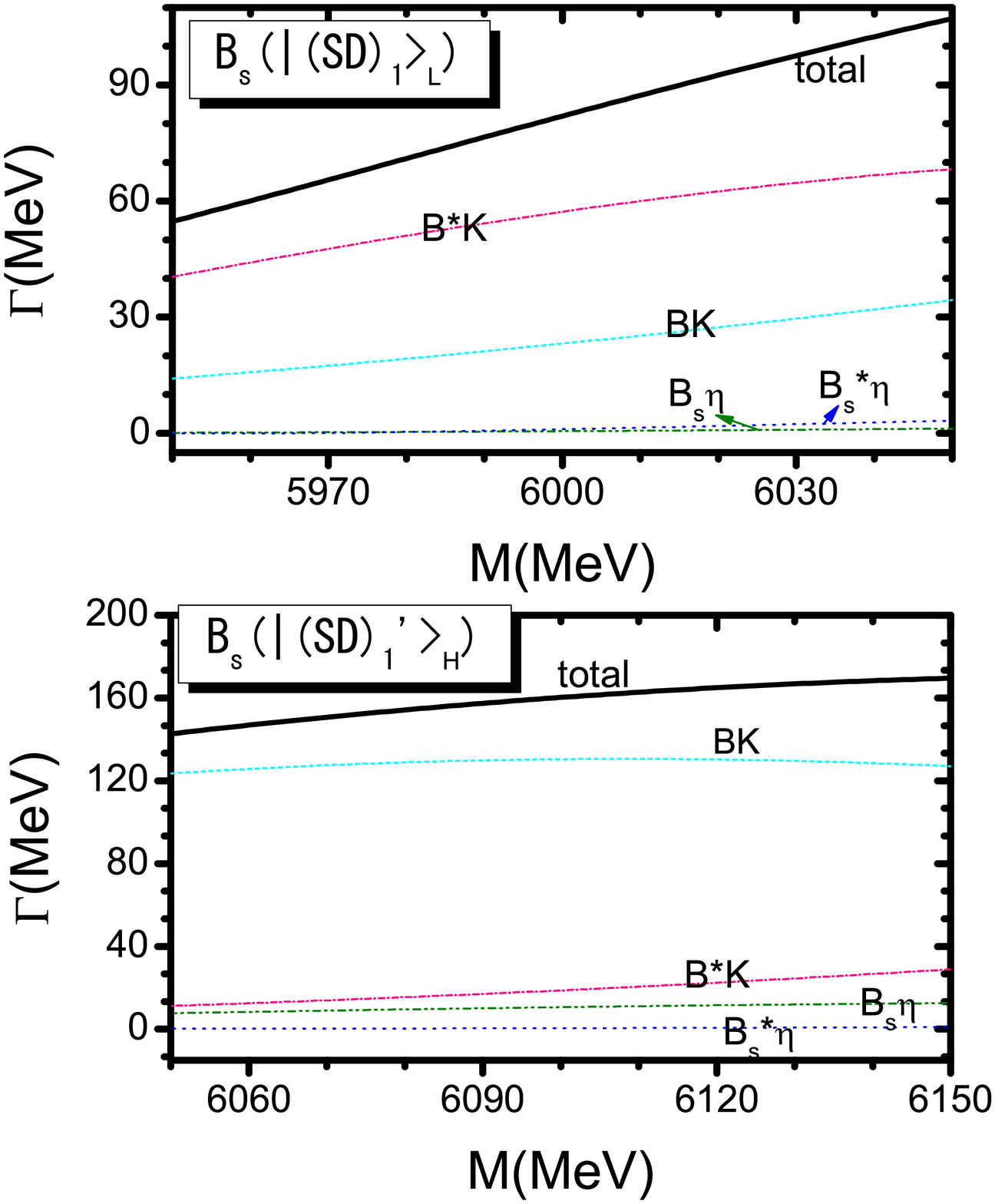}  \caption{The partial decay width and total
decay width for the mixed states via $2^3S_1$-$1^3D_1$ mixing in the
$B$- and $B_s$-meson families as functions of the mass. The mixing
angle is fixed with $\phi=-40^\circ$. In the figure, we have hidden
some decay channels because of their small partial decay
widths.}\label{fig-SDBM}
\end{figure}
\end{center}
\end{widetext}


Firstly, we consider the $2^3S_1$-$1^3D_1$ mixing in the $B$-meson
family. According to the theoretical calculations of the $B$-meson
spectroscopy, the mass of the low-mass state $|(SD)_{1}\rangle_L$ is
$M\sim5.9$ GeV~\cite{Di Pierro:2001uu,Ebert:2009ua}. Thus, we set
the mass of the low-mass mixed state $|(SD)_{1}\rangle_L$ with
$M=5.9$ GeV, and plot its partial decay widths and total decay width
as functions of the mixing angle $\phi$ in Fig.~\ref{fig-SDB}.
Theoretically, the mixing angle $\phi$ between $2^3S_1$ and $1^3D_1$
in the $D$-, $D_s$-, $B$- and $B_s$-meson families should be the
same in the heavy quark symmetry limit. Combining our previous
analysis of the strong decay properties of the $D(2600)$ and
$D_{s1}(2700)$~\cite{Zhong:2010vq,Zhong:2009sk}, the mixing angle
might be $\phi\simeq -(45\pm 16)^\circ$. From Fig.~\ref{fig-SDB}, it
is seen that the mixed state $B(|(SD)_{1}\rangle_L)$ has a width of
$\Gamma\simeq(150\pm50)$ MeV. Its decays are dominated by the $B\pi$
and $B^*\pi$; the partial width ratio between $B\pi$ and $B^*\pi$ is
\begin{eqnarray}
\frac{\Gamma[B\pi]}{\Gamma[B^*\pi]}\simeq 1.0\pm 0.3.
\end{eqnarray}
For the high-mass mixed state $B(|(SD)_{1}'\rangle_H)$, we take its
mass with $M=6.0$ GeV, and plot its partial decay widths and total
decay width as functions of the mixing angle $\phi$ in
Fig.~\ref{fig-SDB} as well, where we can see that this resonance is
a broad state with a width of $\Gamma=(310\pm 90)$ MeV. It decays
mainly through the $B\pi$ and $B_1(5721)\pi$ channels.


Finally, we study the $2^3S_1$-$1^3D_1$ mixing in the $B_s$-meson
family. According to the theoretical predictions in various models,
the masses for the mixed states $|(SD)_{1}\rangle_L$ and
$|(SD)_{1}'\rangle_H$ are around $6.0$ and $6.1$ GeV,
respectively~\cite{Di Pierro:2001uu,Ebert:2009ua}. With these
predicted masses, we plot the partial decay widths and total decay
widths for the mixed states $|(SD)_{1}\rangle_L$ and
$|(SD)_{1}'\rangle_H$ as functions of the mixing angle $\phi$ in
Fig.~\ref{fig-SDB}, respectively. Adopting the mixing angle
$\phi\simeq -(45\pm 16)^\circ$, from Fig.~\ref{fig-SDB} we see that
the low-mass state $|(SD)_{1}\rangle_L$ might have a width of
$\Gamma=(130\pm 50)$ MeV. Its strong decays are dominated by the
$BK$ and $B^*K$ channels. The high-mass state $|(SD)_{1}'\rangle_H$
is a slightly broader state with a width of $\Gamma=(140\pm60)$ MeV.
The decay widths for the $|(SD)_{1}\rangle_L$ and
$|(SD)_{1}'\rangle_H$ are sensitive to the mixing angle.

The predicted masses of these mixed states $|(SD)_1\rangle_L$ and
$|(SD)_{1}'\rangle_H$ in $B$- and $B_s$-meson spectroscopy have a
large uncertainty, which may bring uncertainties to the predicted
decay widths. To investigate this effect, we plot the decay width as
a function of the mass in Fig. \ref{fig-SDBM} with the mixing angle
$\phi=-40^\circ$. It shows that there is an uncertainty of about
$(50\sim 200)$ MeV in the total decay width for the uncertainties of
the mass. The sensitivity of different decay modes to the mass can
also be seen clearly in the plot.

In summary, configuration mixing might exist between $2^3S_1$ and
$1^3D_1$ according to our model prediction. The decay properties are
sensitive to the mixing angle. As flavor partners of the $D(2600)$
and $D_{s1}(2700)$, the low-mass mixed states
$B(|(SD)_{1}\rangle_L)$ and $B_s(|(SD)_{1}\rangle_L)$ have a
relatively narrow width and are most likely to be observed in future
experiments. In the high-mass mixed states, the widths of
charmed-strange state $D_s(|(SD)_{1}'\rangle_H)$ and bottom-strange
state $B_s(|(SD)_{1}'\rangle_H)$ are as comparable as those of the
low-mass mixed states, which are possibly to be found in experiments
as well, while the bottom state $B(|(SD)_{1}'\rangle_H)$ is a broad
state, which might be hard to find in experiments. However, it
should be pointed out that in some
Refs.~\cite{Lu:2014zua,Colangelo:2007ds,Colangelo:2012xi,Godfrey:2013aaa,Wang:2010ydc},
one believes that configuration mixing might be weak between
$2^3S_1$ and $1^3D_1$. Within their models, the strong decay
properties of the $D(2600)$ and/or $D_{s1}(2700)$ can be well
explained by taking them as a pure $2^3S_1$ state. To uncover the
puzzles in the $2^3S_1$ and $1^3D_1$ states, more accurate
measurements for the $D(2600)$ and $D_{s1}(2700)$ as well as a
further search for the missing resonances with $J^P=1^-$ in the
heavy-light meson spectroscopy in experiment is needed.

\subsection{The $1^1D_2$-$1^3D_2$ mixing}

The $D(2750)$ is a new excitation observed by \emph{BABAR}
Collaboration in the $D\pi$ channel with a narrow width of
$\Gamma\simeq71\pm17$ MeV~\cite{delAmoSanchez:2010vq}. This state
might correspond to the $D(2740)$ observed by the LHCb Collaboration
in the same channel recently~\cite{Aaij:2013sza}. Their helicity
analysis indicates that this state should have an unnatural parity,
i.e., $J^P=0^-,1^+,2^-,\cdot\cdot\cdot$. In our previous
work~\cite{Zhong:2010vq}, the new resonance $D(2750)$ was discussed
carefully. We concluded that the spin and parity numbers of the
$D(2750)$ might be $J^P=2^-$, this state is most likely to be the
high-mass mixed state $|1D_2'\rangle_H$ via the $1^1D_2$-$1^3D_2$
mixing:
\begin{equation}\label{mixd}
\left(\begin{array}{c}|1D_2\rangle_L\cr |1D_2'\rangle_H
\end{array}\right)=\left(\begin{array}{cc} \cos\phi_{1D} &\sin\phi_{1D}\cr -\sin\phi_{1D} & \cos\phi_{1D}
\end{array}\right)
\left(\begin{array}{c} |1^1D_2\rangle \cr |1^3D_2\rangle
\end{array}\right),
\end{equation}
with a mixing angle $\phi_{1D}=-(51\pm18)^\circ$. The mixing angle
is consistent with that ($\phi_{1D}=-50.8^\circ$) obtained in the
heavy quark symmetry limit~\cite{Close:2005se}. Our predictions are
in compatible with the recent analysis in Ref.~\cite{Lu:2014zua}. It
should be mentioned that with the mixing angle
$\phi_{1D}=-50.8^\circ$, the low-mass state $|1D_2\rangle_L$ and
high-mass state $|1D_2'\rangle_H$ will correspond to a broad state
and a narrow state, respectively~\cite{Close:2005se,Swanson:2006st}.

The strong decay properties of the low-mass state
$D(|1D_2\rangle_L)$ have been studied in~\cite{Zhong:2010vq} as
well. The mass of $D(|1D_2\rangle_L)$ is likely to be $\sim 2.7$
GeV. This state is a broad state with a width of $\Gamma\simeq
(250\sim 500)$ MeV, and its strong decays are dominated by the
$D^*\pi$ and $D_2(2460)\pi$ channels. The $D(|1D_2\rangle_L)$ might
be hard to observe in experiments because of its broad width.

In our previous study~\cite{Zhong:2009sk}, we also found that the
$1^1D_2$-$1^3D_2$ mixing might be crucial to uncovering the
longstanding puzzles in the $D_{sJ}(2860)$. Many people believe that
the $D_{sJ}(2860)$ might be the $1^3D_3$ state. However, considering
the $D_{sJ}(2860)$ as the $1^3D_3$ state only, one can not
understand the important partial width ratio of
$\Gamma(DK)/\Gamma(D^*K)$ measured by
\emph{BABAR}~\cite{Aubert:2009ah} at all. Considering there are two
$D_s$ states with masses around $2.86$ GeV, one resonance
corresponds to the $1^3D_3$ [denoted by $D_{sJ_1}(2860)$] and the
other resonance is the mixed state $|1D_2\rangle_H$ [denoted by
$D_{sJ_2}(2860)$] via the $1^1D_2$-$1^3D_2$ mixing with the same
mixing angle, we can explain the present strong decay properties of
the $D_{sJ}(2860)$ observed in experiments
naturally~\cite{Zhong:2009sk}. Our conclusion is compatible with the
recent theoretical analysis in Ref.~\cite{Godfrey:2013aaa}.

If the $D_{sJ_2}(2860)$ could be assigned as the high-mass state
$|1D_2'\rangle_H$ indeed, its low-mass partner $|1D_2\rangle_L$
might be observed in experiments as well. The mass of low-mass
partner $|1D_2\rangle_L$ is estimated to be $30\sim 50$ MeV lighter
than that of the high-mass state~\cite{Godfrey:1985xj,Di
Pierro:2001uu,Ebert:2009ua}. Thus, the mass of $|1D_2\rangle_L$
might be in the range of ($2.80\sim2.83$) GeV. Its strong decay
properties have been analyzed in our previous
work~\cite{Zhong:2010vq}. It is found $|1D_2\rangle_L$ is a broad
state with a width of $\Gamma\simeq (260\pm 30)$ MeV. The strong
decays are dominated by the $D^*K$ channel.

It should be emphasized that if both $D(2750)$ and $D_{sJ_2}(2860)$
indeed correspond to the mixed state $|1D_2'\rangle_H$, the
$1^1D_2$-$1^3D_2$ mixing might exist in the $B$- and $B_s$-meson
spectroscopy as well. To provide useful clues for the future
experimental search for these mixed states $|1D_2'\rangle_H$ and
$|1D_2\rangle_L$ in $B$- and $B_s$-meson spectroscopy, we study
their strong decay properties in the following.


\begin{widetext}
\begin{center}
\begin{figure}[ht]
\centering \epsfxsize=6.8 cm \epsfbox{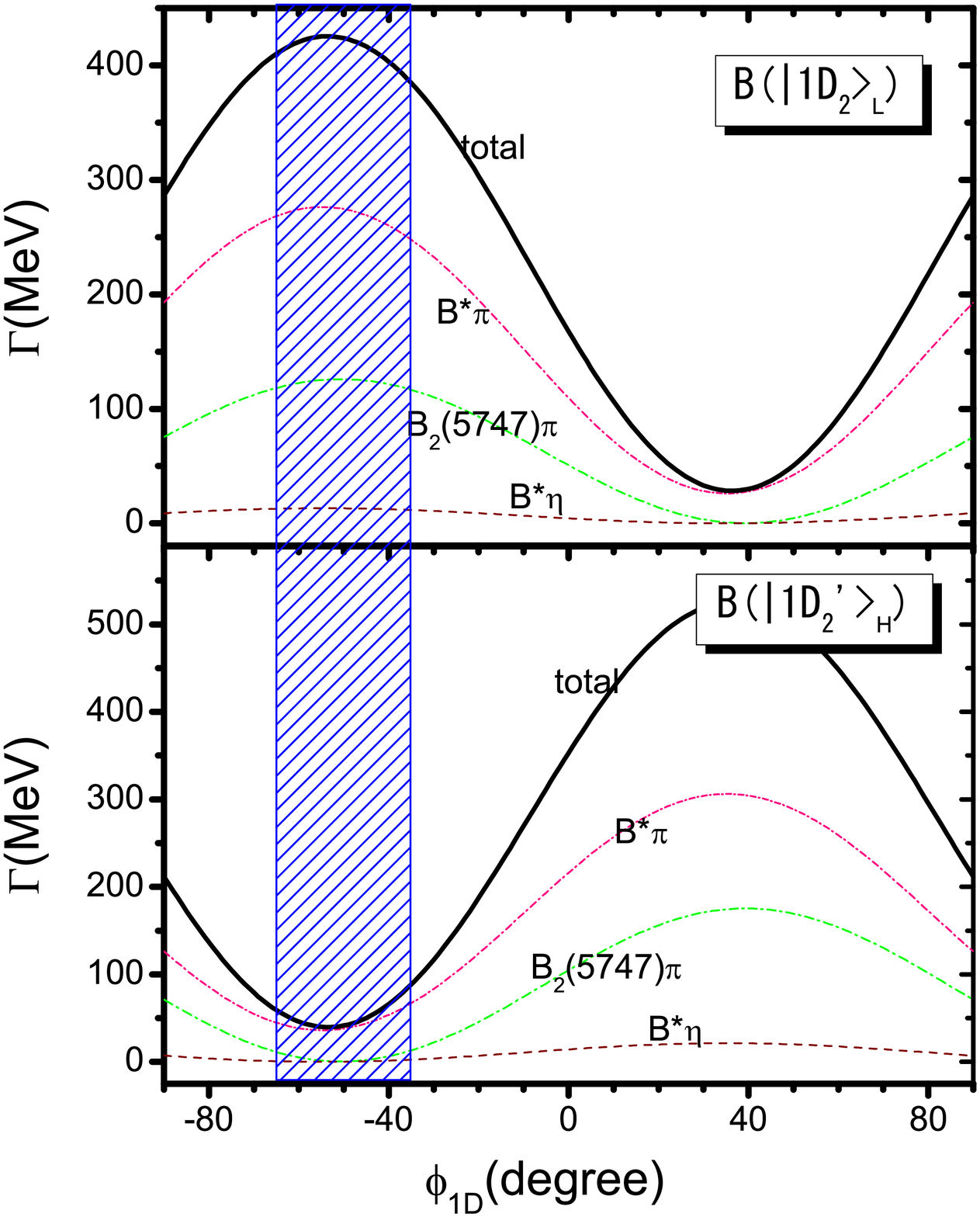}\epsfxsize=6.8 cm
\epsfbox{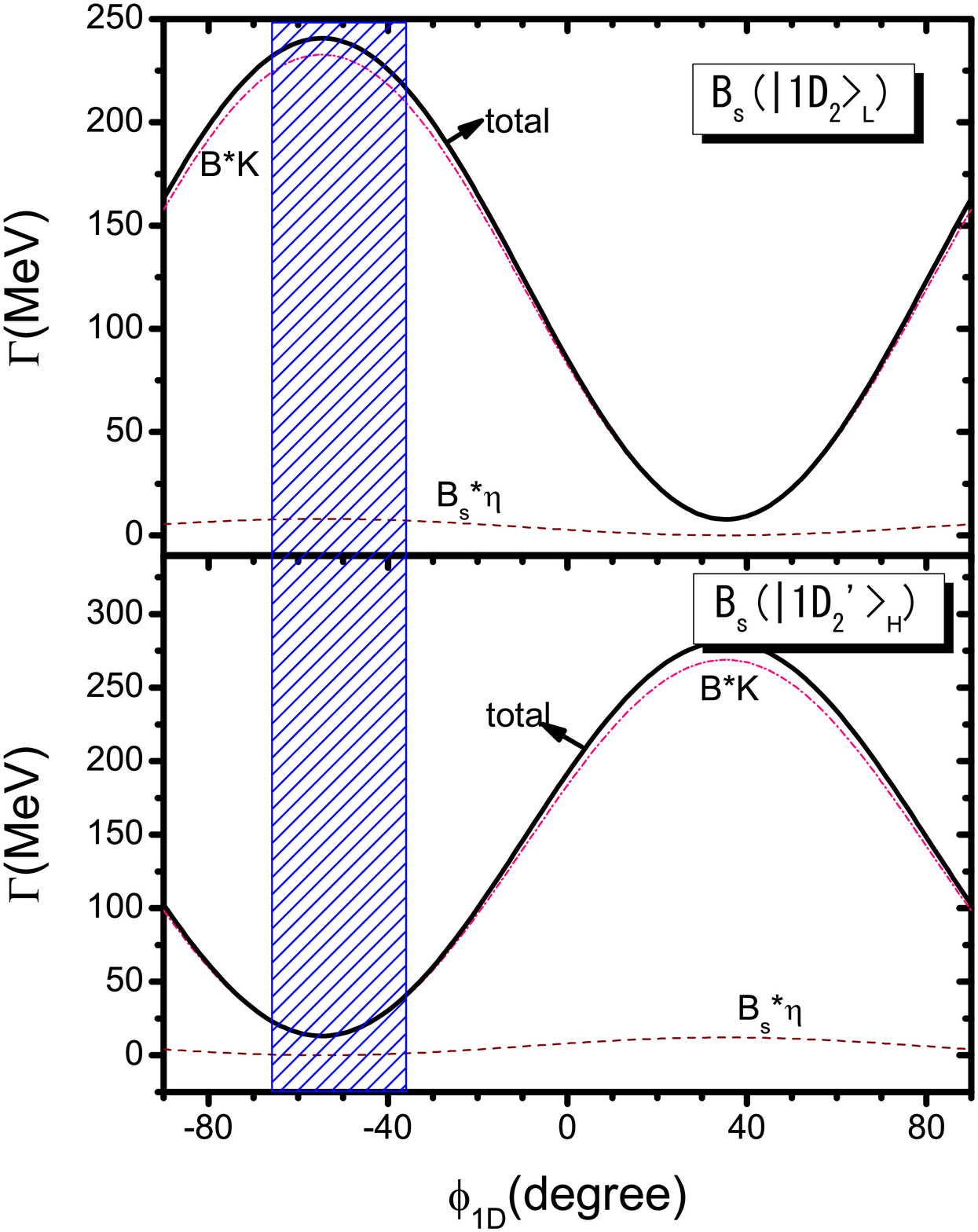}  \caption{The partial decay width and total decay
width for the mixed states via $1^1D_2$-$1^3D_2$ mixing in the $B$-
and $B_s$-meson families as functions of the mixing angle
$\phi_{1D}$. The shaded region corresponds to possible mixing angle
region derived from the strong decay properties of the $D(2750)$.
The masses of the low-mass mixed states $|1D_2\rangle_L$ in the $B$-
and $B_s$-meson families are adopted as 5.95 GeV and 6.05 GeV,
respectively.  The masses of the high-mass mixed states
$|1D_2'\rangle_H$ in the $B$- and $B_s$-meson families are adopted
as 5.98 GeV and 6.08 GeV, respectively. In the figure, we have
hidden some decay channels because of their small partial decay
widths. }\label{fig-DDBH}.
\end{figure}
\end{center}
\end{widetext}

\begin{widetext}
\begin{center}
\begin{figure}[ht]
\centering \epsfxsize=6.5 cm \epsfbox{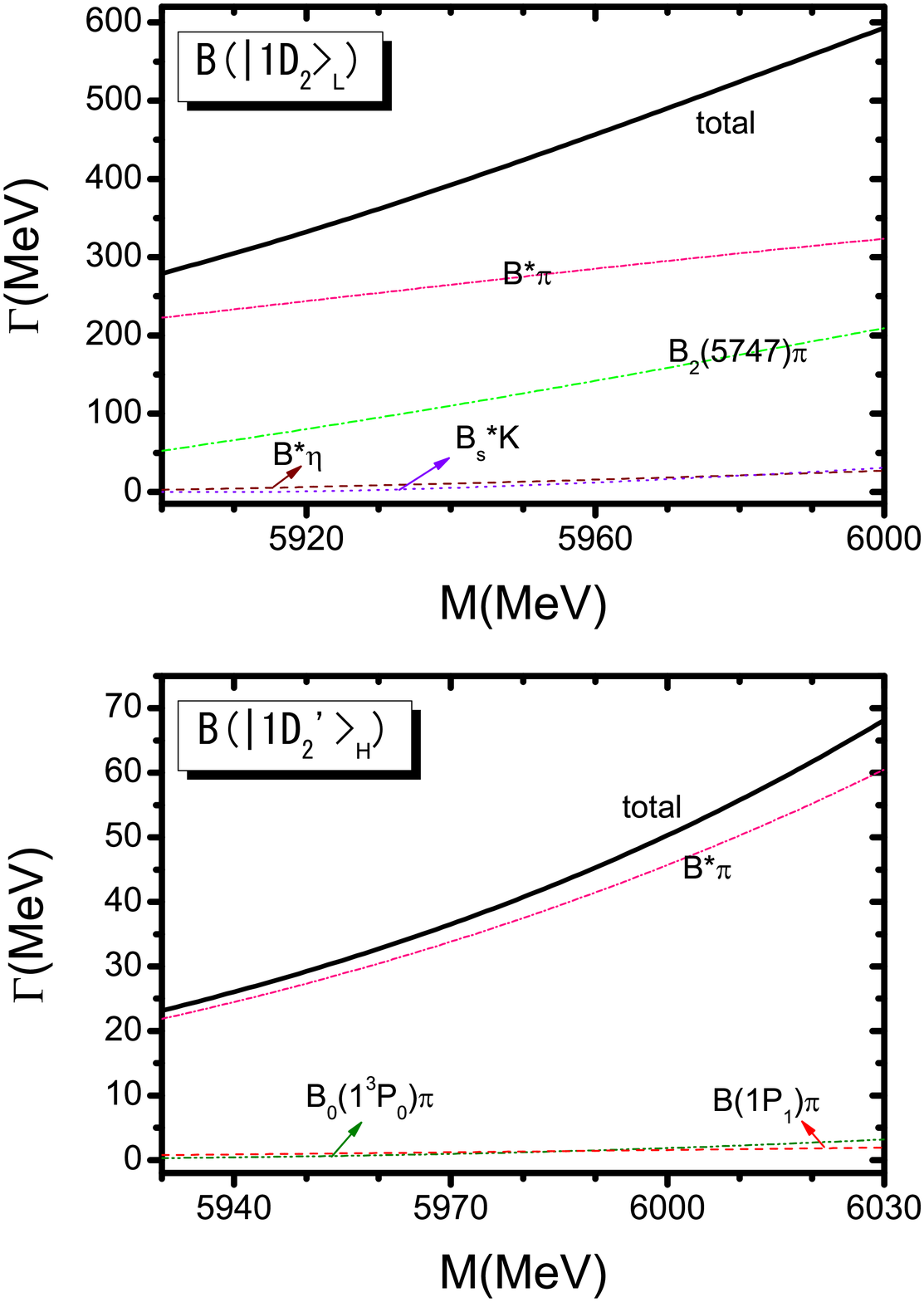} \epsfxsize=6.5 cm
\epsfbox{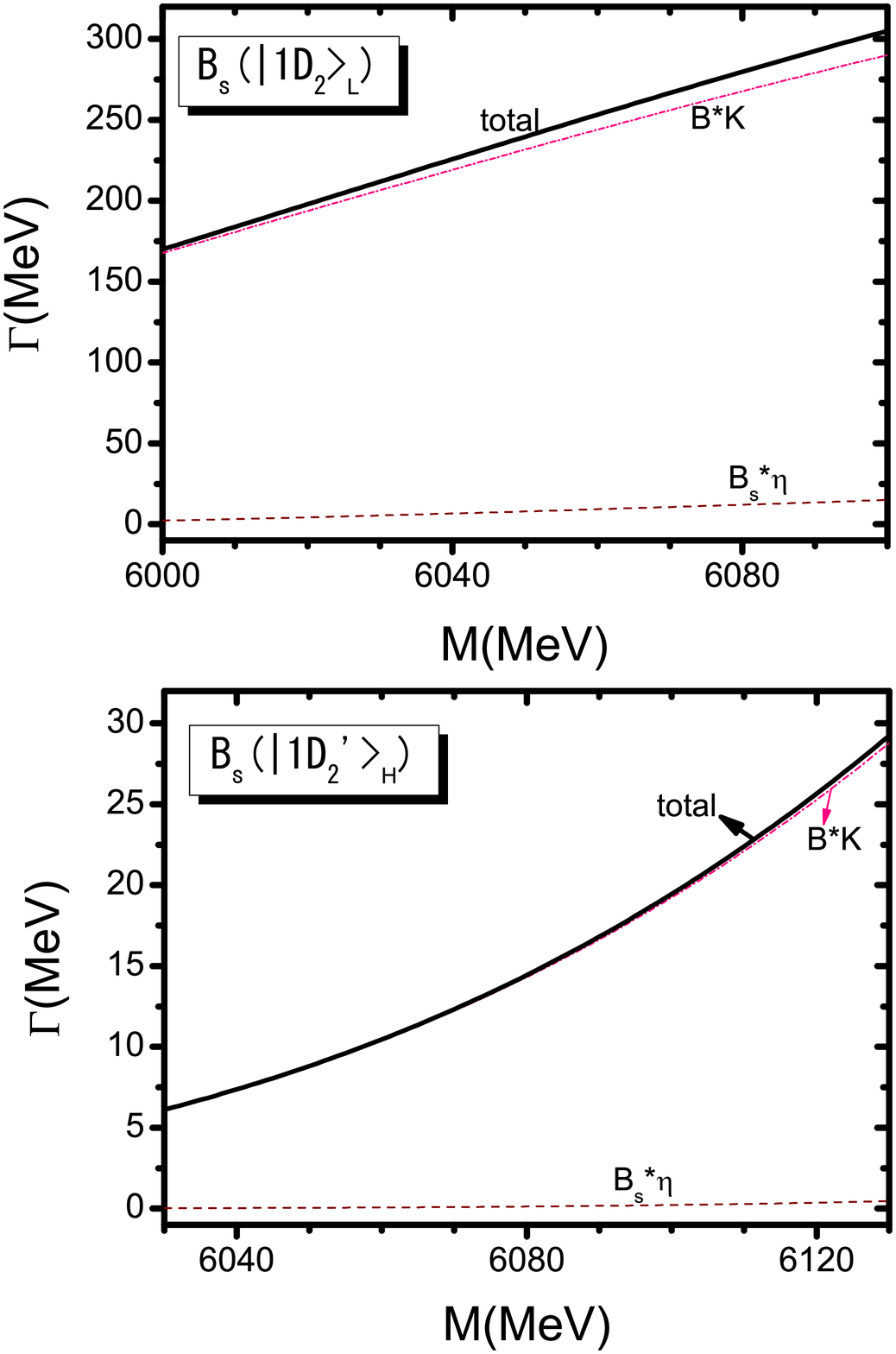} \caption{The partial decay width and total decay
width for the mixed states via $1^1D_2$-$1^3D_2$ mixing in the $B$-
and $B_s$-meson families as functions of the mass. The mixing angle
is adopted as $\phi_{1D}=-50.8^\circ$. In the figure, we have hidden
some decay channels because of their small partial decay widths.
}\label{fig-DDBHM}.
\end{figure}
\end{center}
\end{widetext}

Firstly, we study the $1^1D_2$-$1^3D_2$ mixing in the $B$-meson
family. According to the predictions in
Refs.~\cite{Godfrey:1985xj,Di Pierro:2001uu,Ebert:2009ua}, the mass
of the high-mass mixed state $|1D_2'\rangle_H$ is in the range of
$(6.04\sim 6.12)$ GeV. However, we find the masses of the
$|1D_2'\rangle_H$ resonances might be systemically overestimated by
$\sim 100$ MeV in theory, if the $D(2750)$ and $D_{sJ_2}(2860)$
correspond to the mixed state $|1D_2'\rangle_H$. Thus, the mass of
the $|1D_2'\rangle_H$ in the $B$-meson family might be
$M\simeq(5.94\sim 6.02)$ GeV.

Taking the mass of $B(|1D_2'\rangle_H)$ with $M=5.98$ GeV, we plot
the partial decay widths and total decay width as functions of the
mixing angle $\phi_{1D}$ in Fig.~\ref{fig-DDBH}. It is shown that
the decay width of the high-mass state $|1D_2'\rangle_H$ is about
$\Gamma\simeq(40 \sim 90)$ MeV, when we take the mixing angle with
$\phi_{1D}=(-51\pm18)^\circ$. The main decay channels are $B^*\pi$
and $B_2(5747)\pi$. The predicted branching ratio is
\begin{eqnarray}
\frac{\Gamma[B^*\pi]}{\Gamma_{\mathbf{total}}}\simeq 70\%\sim 93\%.
\end{eqnarray}
The partial widths for the $B(1^3P_0)\pi$, $B(1P_1)\pi$,
$B(1P_1')\pi$, $B^*\eta$ and $B_s^{*}K$ channels are tiny and are
not shown in the figure. Such a narrow state, $B(|1D_2'\rangle_H)$,
is most likely to be observed in the $B^*\pi$ channel.

For the low-mass mixed state $|1D_2\rangle_L$, its mass might be
$(20\sim 50)$ MeV lighter than that of the
$|1D_2\rangle_H$~\cite{Godfrey:1985xj,Di
Pierro:2001uu,Ebert:2009ua}. Thus, the mass of $|1D_2\rangle_L$
might be in the range $M=(5.9\sim6.0)$ GeV. The dependence of the
strong decay width of the $|1D_2\rangle_L$ on the mass is shown in
Fig.~\ref{fig-DDBHM} as well. From the figure, we can see that the
low-mass state $B(|1D_2\rangle_L)$ is a very broad state with a
width of $\Gamma\simeq (400\pm 20)$ MeV. Its strong decays are
governed by the $B^*\pi$ and $B_2(5747)\pi$ channels. This broad
state might be hard to observe in experiments.


Secondly, we study the $1^1D_2$-$1^3D_2$ mixing in the $B_s$-meson
family. According to the theoretical predictions, the mass gap
between the $B_s$ resonances and the $B$ resonances is about 100
MeV~\cite{Godfrey:1985xj,Di Pierro:2001uu,Ebert:2009ua}. Thus, the
high-mass state $B_s(|1D_2'\rangle_H)$ might have a mass of $M\simeq
(6.04\sim 6.12)$ GeV, while the mass of the low-mass state
$B_s(|1D_2\rangle_L)$ might be $M\simeq (5.99\sim 6.10)$ GeV.

Taking the mass of the $B_s(|1D_2'\rangle_H)$ with $M=6.08$ GeV, we
plot the  partial widths and total decay width as functions of the
mixing angle in Fig.~\ref{fig-DDBH}. Adopting the physical mixing
angle $\phi_{1D}=(-51\pm18)^\circ$ determined before, we find the
$B_s(|1D_2'\rangle_H)$ is a fairly narrow state with a width of
$\Gamma\simeq (13\sim 50)$ MeV. Its decays are dominated by the
$B^*K$ channel. The predicted branching ratio is
\begin{eqnarray}
\frac{\Gamma[B^*K]}{\Gamma_{\mathbf{total}}}\simeq 100\%.
\end{eqnarray}
This narrow state is most likely to be discovered in
the $B^*K$ channel in future experiments.

Furthermore, we analyze the strong decay properties of the low-mass
state $B_s(|1D_2\rangle_L)$. Its partial decay widths and total
width as functions of the mixing angle are plotted in
Fig.~\ref{fig-DDBH} as well, where we have fixed the mass with
$M=6.05$ GeV. From the figure, we find that the low-mass state
$B_s(|1D_2\rangle_L)$ has a broad width of $\Gamma\simeq 240$ MeV,
and its decays are governed by the $B^*K$ channel.

Finally, we study the effects of the mass uncertainties of the
mixing states $|1D_2\rangle_L$ and $|1D_{2}'\rangle_H$ in $B$ and
$B_s$ spectroscopy on the decay widths. To investigate this effect,
we plot the decay width as a function of the mass in Fig.
\ref{fig-DDBHM} with the mixing angle fixed at
$\phi_{1D}=-50.8^\circ$. The sensitivity of different decay modes to
the mass can also be seen clearly in the plot.

As a whole, the high-mass mixed states $B(|1D_2'\rangle_H)$ and
$B_s(|1D_2'\rangle_H)$ might be narrow states, which might be
observed in the $B^*\pi$ and $B^*K$ channels, respectively. However,
the low-mass mixed state $|1D_2\rangle_L$ should be a broad state.
The typical total decay widths for the low-mass mixed states
$D(|1D_2\rangle_L)$ and $B(|1D_2\rangle_L)$ are usually larger than
$300$ MeV, which might be too broad to be observed in experiments.
However, those low-mass mixed states containing a strange quark,
$D_s(|1D_2\rangle_L)$ and $B_s(|1D_2\rangle_L)$, have a relatively
narrower width, $\sim 250$ MeV, which have some possibilities to be
observed in future experiments.

\subsection{The $1^3D_3$ states}


The $D(2760)$ is a new excitation in the $D$-meson family observed
by \emph{BABAR} in the $D\pi$ channel with a mass of
$M=2763.3\pm4.6$ MeV and a width of $\Gamma=60.9\pm 8.7$ MeV.
Although its mass and width are very close to those of the
$D(2750)$, they might be two different resonances due to several
reasons, which have been explained in Ref.~\cite{Zhong:2010vq}.
Recently, this state was confirmed by the LHCb Collaboration in the
$D\pi$ and $D^*\pi$ channels, and their helicity analysis indicates
that this state should have a natural parity, i.e.,
$J^P=0^+,1^-,2^+,3^-,\cdot\cdot\cdot$. In our previous
work~\cite{Zhong:2010vq}, this new resonance $D(2760)$ was studied
carefully. It is predicted that the $D(2760)$ could be assigned to
the $1^3D_3(J^P=3^-)$ state. Our conclusion is consistent with that
in Refs.~\cite{Wang:2010ydc,Sun:2010pg,Lu:2014zua}. Taking the
$D(2760)$ as the assignment of the $1^3D_3$, the total decay width
is $\Gamma\simeq68$ MeV, which is consistent with the data. Its
strong decays are dominated by the $D\pi$ and $D^*\pi$ channels, and
the partial width ratio is
\begin{eqnarray}
\frac{\Gamma[D^*\pi]}{\Gamma[D\pi]}\simeq 0.65,
\end{eqnarray}
which can be tested in future experiments.

\begin{figure}[ht]
\centering \epsfxsize=7 cm \epsfbox{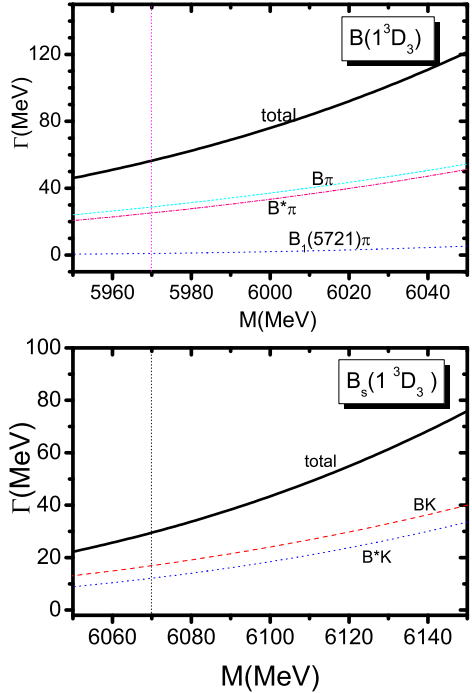} \caption{The partial
decay widths and total decay width of the $1^3D_3$ excitations in
the $B$- and $B_s$-meson families as functions of the mass. In the
figure, we have hidden some decay channels because of their small
partial decay widths. }\label{fig-DFORB}
\end{figure}


Furthermore, in our previous work~\cite{Zhong:2009sk}, we suggested
there are two largely overlapping resonances at 2.86 GeV in the
$D_s$-meson family. One state corresponds to the $1^3D_3$
$(J^P=3^-)$ state [denoted by $D_{sJ_1}(2860)$], which dominantly
decays into the $DK$ channel, while the other one corresponds to the
mixed state $|1D_2'\rangle_H$ with $J^P=2^-$ [denoted by
$D_{sJ_2}(2860)$], which dominantly decays into the $D^*K$ channel.
Considering the $D_{sJ_1}(2860)$ as the $1^3D_3$ state, we have
studied its strong decay properties. The predicted decay width is
$\Gamma\simeq 45$ MeV, which is highly comparable with the
experimental data ($\Gamma=58\pm11$ MeV). This state has two main
decay channels $DK$ and $D^*K$, and the predicted partial width
ratio is
\begin{eqnarray}
\frac{\Gamma[D^*K]}{\Gamma[DK]}\simeq 0.5,
\end{eqnarray}
which can be tested in future experiments. The present experimental
measurement of the ratio,
$\Gamma[D^*K]/\Gamma[DK]=1.10\pm0.15\pm0.19$, might include the
contributions of both $D_{sJ_1}(2860)$ and $D_{sJ_2}(2860)$. Thus,
its value is obviously larger than the theoretical predictions by
assuming the $D_{sJ}(2860)$ as the $1^3D_3$ state only. If both
$D(2760)$ and $D_{sJ_1}(2860)$ correspond to the $1^3D_3$ excitation
indeed, their flavor partners in $B$ and $B_s$ spectroscopy should
be observed in future experiments as well.


In the $B$-meson family, the predicted mass of $1^3D_3$ is around
$6.0\sim 6.1$ GeV~\cite{Godfrey:1985xj,Di
Pierro:2001uu,Ebert:2009ua,Sun:2014wea}. Considering the $D(2760)$
and $D_{sJ_1}(2860)$ as the $1^3D_3$ assignment, their mass is
systemically overestimated by $50\sim 100$ MeV in theory. Thus, we
estimate the mass of $B(1^3D_3)$ is in the range of $(5.95\sim
6.05)$ GeV. We calculate the strong decay properties of $B(1^3D_3)$,
which are shown in Fig.~\ref{fig-DFORB}. From the figure, it is seen
that the $B(1^3D_3)$ is a narrow state with a width of $\Gamma\simeq
50\sim 120$ MeV. Its decays are dominated by the $B\pi$ and $B^*\pi$
channels. The partial width ratio between $B\pi$ and $B^*\pi$ is
insensitive to the mass, and its predicted value is
\begin{eqnarray}
\frac{\Gamma[B^*\pi]}{\Gamma[B\pi]}\simeq 0.9.
\end{eqnarray}

Recently, the CDF Collaboration claimed that they found some
evidence of a new resonance $B(5970)$ in the $B\pi$
channel~\cite{Aaltonen:2013atp}. The central value of the width is
$\Gamma\sim 70$ MeV. From Fig.~\ref{fig-DFORB}, it is interestingly
found that the $B(5970)$ is most likely to be the $1^3D_3$
assignment. Considering the $B(5970)$ as the $1^3D_3$ state, the
predicted decay width is $\Gamma\simeq 60$ MeV, and the $B\pi$ and
$B^*\pi$ decay channels are its dominant decay modes. Our
predictions are in a good agreement with the observations. Although
the mass of the $B(5970)$ is close to the estimated masses of the
mixed states $B(|(SD)_{1}\rangle_L)$ and $B(|1D_2'\rangle_H)$, they
are not good candidates for the $B(5970)$. For the
$B(|(SD)_{1}\rangle_L)$, when we set its mass with $\sim 5.97$ GeV,
its theoretical width is $\sim300$ MeV, which is too large to
compared with the width of the $B(5970)$, while for the
$B(|1D_2'\rangle_H)$, its strong decays are governed by the
$B^*\pi$, and the $B\pi$ channel is forbidden. Thus, it could not be
assigned to the $B(5970)$.


Finally, we study the strong decays of $1^3D_3$ in the $B_s$-meson
family. Usually, the mass of the $B_s$ resonances is about 100 MeV
larger than that of the $B$ resonances~\cite{Godfrey:1985xj,Di
Pierro:2001uu,Ebert:2009ua}. Thus, we estimate the mass of
$B_s(1^3D_3)$ might be in the range of $(6.05\sim 6.15)$ GeV. We
have studied the strong decay properties of $B_s(1^3D_3)$, which are
shown in Fig.~\ref{fig-DFORB}. It is found that the $B_s(1^3D_3)$
might be a narrow state with a width of $\Gamma\simeq (25\sim 75)$
MeV, and its strong decays are dominated by the $BK$ and $B^*K$
channels. If the $B(5970)$ corresponds to the $B(1^3D_3)$ indeed,
the mass of the $B_s(1^3D_3)$ might be $M\simeq6.07$ GeV. With this
mass, the predicted width for the $B_s(1^3D_3)$ is
\begin{eqnarray}
\Gamma\simeq 30\ \mathrm{MeV},
\end{eqnarray}
and the predicted ratio between $BK$ and $B^*K$ is
\begin{eqnarray}
\frac{\Gamma[BK]}{\Gamma[B^*K]}\simeq 1.4.
\end{eqnarray}
We hope the experimenters can carry out a search for the
$B_s(1^3D_3)$ in the $BK$ and $B^*K$ channels, which is also helpful
to clarify the nature of the newly observed state $B(5970)$.

In brief, the $D(2760)$, $D_{sJ_1}(2860)$ and $B(5970)$ might be
classified as the low-lying $D$-wave excitations with $J^P=3^-$
(i.e., $1^3D_3$). The last unobserved one in the $B_s$-meson family
should be a narrow state, which is most likely to be found in the
$BK$ and $B^*K$ channels.

\begin{widetext}
\begin{center}
\begin{figure}[ht]
\centering \epsfxsize=15.0 cm \epsfbox{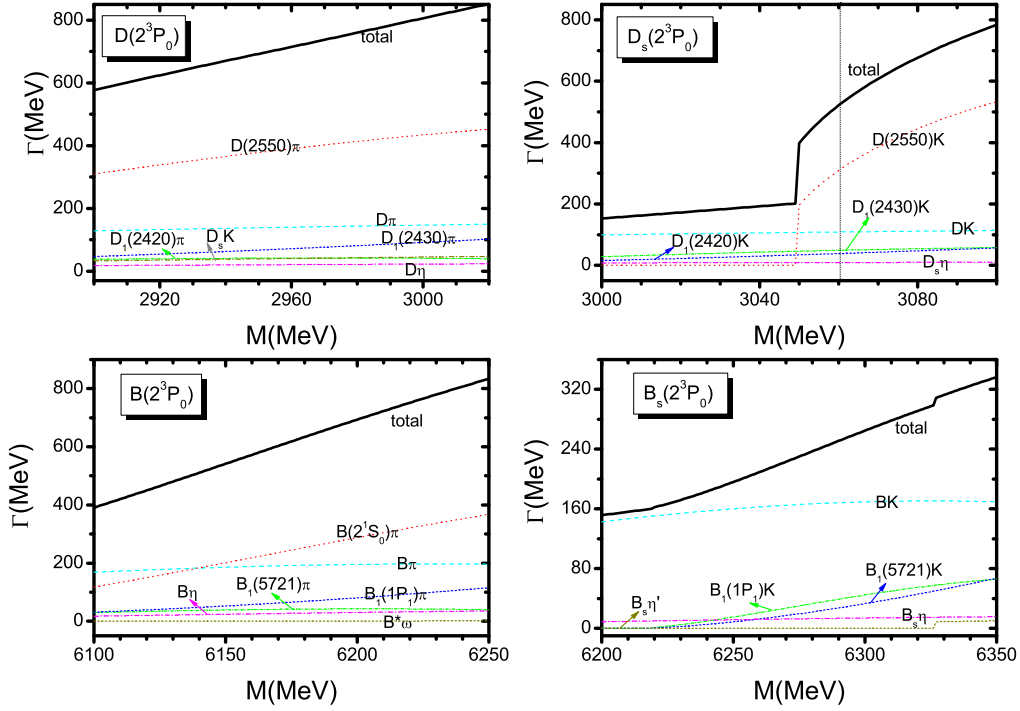} \caption{ The
partial decay width and total decay widths of the $2^3P_{0}$ states
in the heavy-light mesons as functions of the mass. In the figure,
we have hidden some decay channels because of their small partial
decay widths. The mass of the $B(2^1S_0)$ is adopted as $M=5886$
MeV.}\label{fig-3P0}
\end{figure}
\end{center}
\end{widetext}

\subsection{The $2^3P_0$ states}


In the $D$-meson family, the predicted mass for the $2^3P_0$ state
is $\sim 2.95$ GeV~\cite{Di Pierro:2001uu,
Ebert:2009ua,Sun:2013qca}. For the uncertainties of the predicted
mass, we vary the mass of $D(2^3P_0)$ in the range of $M=(2.9\sim
3.0)$ GeV, and plot the strong decay properties in
Fig.~\ref{fig-3P0}. From the figure it is seen that the $D(2^3P_0)$
might be a very broad state with a width of $\Gamma\simeq 600\sim
800$ MeV. This state dominantly decays into the $D(2550)\pi$, $D\pi$
and $D(2430)\pi$ channels. A very broad width for the $D(2^3P_0)$
was also predicted in a recent study with the $^3P_0$
model~\cite{Lu:2014zua}. Such a broad resonance might be difficult
to observe in experiments.

Recently, the LHCb Collaboration observed a new excited $D$-meson
state $D_J^*(3000)$ in the $D\pi$ channel with a rather narrow
width, $\Gamma\simeq 110$ MeV~\cite{Aaij:2013sza}. Their further
helicity analysis indicates that this new state should have a
natural parity, that is, the spin-parity number of the $D_J^*(3000)$
should be $J^P=0^+, 1^-, 2^+, \cdot\cdot\cdot$. In
Ref.~\cite{Sun:2013qca}, Sun \emph{et al.} believed that the
$D_J^*(3000)$ could be explained as the $D(2^3P_0)$. If we exclude
the contribution of the $D(2550)\pi$ channel to the decay width, our
results agree with the predictions in Ref.~\cite{Sun:2013qca}.
However, including the contribution of the $D(2550)\pi$ decay
channel, our calculations show that the decay width, $\Gamma\simeq
800$ MeV, is too broad to compare with the data. Thus, the
$D_J^*(3000)$ as the $D(2^3P_0)$ assignment still bears
controversies, although the parity and decay modes are consistent
with observations.


In the $D_s$-meson family, the predicted mass of the $2^3P_0$ is
about $3.05$ GeV, which is $\sim100$ MeV heavier than that of
$D(2^3P_0)$~\cite{Di Pierro:2001uu, Ebert:2009ua}. Considering the
uncertainties of the predicted mass, we plot the strong decay
properties of $D_s(2^3P_0)$ as functions of the mass in
Fig.~\ref{fig-3P0}. It is shown that if the mass of $D_s(2^3P_0)$ is
less than $3.05$ GeV, the $D_s(2^3P_0)$ has a moderate width of
$\Gamma\simeq 150\sim 200$ MeV, and its strong decays are governed
by the $DK$ channel. In this case, the $D_s(2^3P_0)$ is most likely
to be observed in the $DK$ channel. On the other hand, if the mass
of $D_s(2^3P_0)$ is larger than $3.05$ GeV, the decay channel
$D(2550)K$ will be opened, which dominates the strong decays. Then,
the $D_s(2^3P_0)$ might be a broad state with a width of
$\Gamma\simeq 400\sim 800$ MeV. In this case, the $D_s(2^3P_0)$
might be too broad to observe in experiments.


In the $B$-meson family, the predicted mass of the $2^3P_0$ is about
$6.2$ GeV~\cite{Di Pierro:2001uu, Ebert:2009ua,Sun:2014wea}. To know
about the decay properties of the $B(2^3P_0)$, in Fig.~\ref{fig-3P0}
we plot its decay width as a function of mass in the range of
$M=(6.1\sim6.25)$ GeV. From the figure it is seen that the
$B(2^3P_0)$ might be a broad state with a width of $\Gamma\simeq
400\sim 800$ MeV. This state dominantly decays into the
$B(1^3P_0)\pi$, $B\pi$, and $B(1P_1)\pi$ channels. Such a broad
resonance might be difficult to observe in experiments.


In the $B_s$-meson family, the predicted mass of the $2^3P_0$ is
about $M\simeq 6.3$ GeV~\cite{Di Pierro:2001uu,
Ebert:2009ua,Sun:2014wea}. To know about the decay properties of the
$B_s(2^3P_0)$, in Fig.~\ref{fig-3P0} we plot its decay width as a
function of mass in the range of $M=(6.2\sim6.35)$ GeV. From the
figure it is seen that the decay width of $B_s(2^3P_0)$ is
$\Gamma\simeq 240$ MeV with $M\simeq 6.27$ GeV, which is much
narrower than that of $B(2^3P_0)$. The strong decays of
$B_s(2^3P_0)$ are governed by the $BK$ channel. The other decay
modes $B(1P_1)K$ and $B(1P_1')K$ also contribute to the decays
obviously, if the mass of $B_s(2^3P_0)$ is larger than $6.25$ GeV.
Since the predicted width of $B_s(2^3P_0)$ is not very broad, this
resonance might be observed in the $BK$ channel in future
experiments.

As a whole, in the $2^3P_0$ states, the widths of $D(2^3P_0)$ and
$B(2^3P_0)$ might be too broad to observe in experiments. However,
the widths of the resonances $D_s(2^3P_0)$ and $B_s(2^3P_0)$ are
relatively narrower, thus, they might be observed in the $DK$ and
$BK$ channels, respectively.

\subsection{The $2^1P_1$-$2^3P_1$ mixing}


The $D_{sJ}(3040)$ is a new broad state in the $D_s$-meson family
observed by $BABAR$ in the $D^*K$ channel~\cite{Aubert:2009ah}. It
has a mass of $M=3044\pm8^{+30}_{-5}$ MeV and a broad width of
$\Gamma=239\pm35^{+46}_{-42}$ MeV. According to our previous
study~\cite{Zhong:2009sk}, the $D_{sJ}(3040)$ can be naturally
explained as the low-mass mixed state $|2P_{1}\rangle_L$ via the
$2^1P_1$-$2^3P_1$ mixing:
\begin{equation}\label{mixp}
\left(\begin{array}{c}|2P_{1}\rangle_L\cr |2P_{1}'\rangle_H
\end{array}\right)=\left(\begin{array}{cc} \cos\phi_{2P} &\sin\phi_{2P}\cr -\sin\phi_{2P} & \cos\phi_{2P}
\end{array}\right)
\left(\begin{array}{c} |2^1P_1\rangle \cr |2^3P_1\rangle
\end{array}\right),
\end{equation}
with a mixing angle $\phi_{2P}\simeq-(51\pm27)^\circ$. This mixing
angle is consistent with that ($\phi_{2P}=-54.7^\circ$) derived in
the heavy quark symmetry limit~\cite{Close:2005se}. It should be
mentioned that with the mixing angle $\phi_{2P}=-54.7^\circ$, the
low-mass state $|2P_{1}\rangle_L$ usually has a broad width, while
the high-mass state $|2P_{1}'\rangle_H$ has a narrower
width~\cite{Close:2005se,Swanson:2006st}. As a broad mixed state,
the strong decays of the $D_{sJ}(3040)$ are dominated by the $D^*K$
channel. The partial widths of $D_0(2400)K$, $D_1(2430)K$, and
$D_2(2460)K$ are sizable as well. Our conclusion is consistent with
that obtained in Refs.~\cite{Sun:2009tg,Ebert:2009ua,Mohler:2011ke}.

As the physical partner of $|2P_{1}\rangle_L$, the high-mass state
$|2P_{1}'\rangle_H$ might be observed in experiments. Assuming the
mass of the $|2P_{1}'\rangle_H$ is in the range of $(3.04\sim3.2)$
GeV, its strong decay properties have been analyzed in our previous
work~\cite{Zhong:2009sk}. The decay width of $|2P_{1}'\rangle_H$ is
narrower than that of the $D_{sJ}(3040)$, which increases fast with
the increasing mass. The main decay channels might be $D_0(2400)K$,
$D_1(2430)K$ and $DK^*$.

If the $D_{sJ}(3040)$ could indeed be assigned as a mixed state via
the $2^1P_1$-$2^3P_1$ mixing, the other mixed states between
$2^1P_1$ and $2^3P_1$ in the $D$-, $B$- and $B_s$-meson spectroscopy
might be found in future experiments.

\begin{figure}[ht]
\epsfxsize=6.5 cm \epsfbox{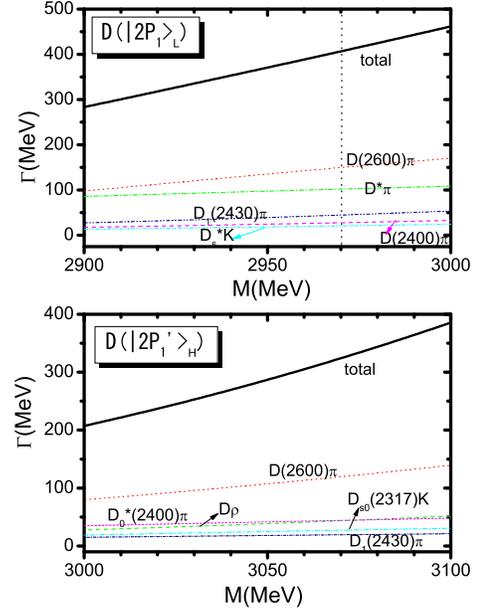} \caption{ The partial decay
widths and total decay width of the mixed states via the
$2^1P_1$-$2^3P_1$ mixing in the $D$-meson family as functions of the
mass. Here, the mixing angle $\phi_{2P}=-54.7^\circ$ is adopted. In
the figure, we have hidden some decay channels because of their
small partial decay widths. }\label{fig-ppfordm}
\end{figure}

Firstly, we study the mixed states of $2^1P_1$-$2^3P_1$ in the
$D$-meson family. According to the predictions in
Refs.~\cite{Godfrey:1985xj,Di Pierro:2001uu,Ebert:2009ua}, the mass
of the low-mass state $D(|2P_1\rangle_L)$ might be in the range of
$(2.9\sim 3.0)$ GeV, while the mass of the high-mass state
$D(|2P_1'\rangle_H)$ might be in the range of $(3.0\sim 3.1)$ GeV.
Adopting the mixing angle $\phi_{2P}=-54.7^\circ$ predicted in the
heavy quark symmetry limit, we have plotted their partial decay
widths and total decay width as functions of the mass in
Fig.~\ref{fig-ppfordm}. From the figure, it is seen that the
low-mass state $D(|2P_1\rangle_L)$ is a broad state with a width of
$\Gamma\simeq (370\pm 90)$ MeV, its strong decays are dominated by
the $D^*\pi$ and $D(2600)\pi$ channels, and the partial decay widths
of the $D_0(2400)\pi$, $D_1(2430)\pi$ and $D^*_sK$ channels are
sizable as well.

Although the high-mass state $D(|2P_1'\rangle_H)$ has a relatively
narrower width than the low-mass state $D(|2P_1\rangle_L)$, the
$D(|2P_1'\rangle_H)$ is also a very broad state with a width of
$\Gamma\simeq (300\pm 90)$ MeV when we adopt its mass in the range
of $(3.0\sim 3.1)$ GeV. Its strong decays are governed by the
$D(2600)\pi$ channel. The other channels $D\rho$, $D_0(2400)\pi$,
$D_{s0}(2317)K$ and $D_1(2430)\pi$ also have sizable contributions
to the decays.

\begin{figure}[ht]
\centering \epsfxsize=8 cm \epsfbox{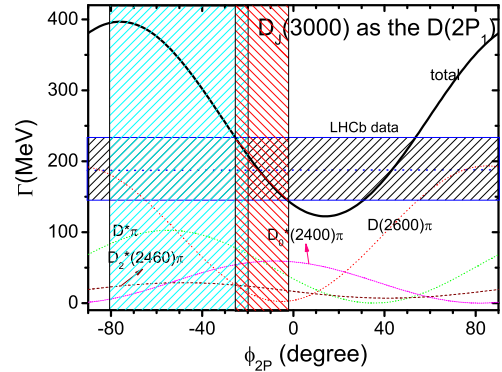} \caption{ The
variation the decay properties of the mixed state $|2P_{1}\rangle_L$
as a candidate of $D_J(3000)$ with the mixing angle $\phi_{2P}$. The
left shaded band corresponds to the possible mixing angle region
derived from the strong decay properties of the $D_{sJ}(3040)$,
while the right shaded band corresponds to the mixing angle region
derived from the strong decay properties of the $D_{J}(3000)$. In
the figure, we have hidden some decay channels for their small
partial decay widths. }\label{fig-ppdl}
\end{figure}

Recently, a new resonance $D_J(3000)$ was observed by the LHCb. Its
measured mass and width are $M\simeq 2972$ MeV and $\Gamma\simeq
188\pm45$ MeV, respectively. Furthermore, the helicity analysis from
LHCb indicates that this state has an unnatural parity, that is, its
spin-parity should be $J^P=0^-, 1^+, 2^-, \cdot\cdot\cdot$.
According to the observed mass, the observed $D^*\pi$ decay mode and
possible $J^P$ numbers of the $D_J(3000)$, we predict that the
$D_J(3000)$ might be a good candidate for the low-mass state
$D(|2P_1\rangle_L)$.  As the $D(|2P_1\rangle_L)$ candidate with the
mixing angle $\phi_{2P}=-54.7^\circ$ derived in the heavy quark
symmetry limit, the $D_J(3000)$ might have a broad width of
$\Gamma\simeq 360$ MeV (see Fig.~\ref{fig-ppfordm}), which is about
a factor 2 larger than the data ($\Gamma\simeq 188\pm45$ MeV). The
predicted partial decay width ratio of the main two decay channels
$D^*\pi$ and $D(2600)\pi$ is
\begin{eqnarray}
\frac{\Gamma[D^*\pi]}{\Gamma[D(2600)\pi]}\simeq 0.7.
\end{eqnarray}

However, we have noticed that the mixing angle $\phi_{2P}$
determined by the strong decay properties of the $D_{sJ}(3040)$
bares a large uncertainty. Taking the $D_J(3000)$ as a candidate of
the $D(|2P_1\rangle_L)$ state, we have plotted the partial widths
and total decay width as functions of the mixing angle $\phi_{2P}$
in Fig.~\ref{fig-ppdl}. From the figure, it is found that the decay
width is very sensitive to the mixing angle. If we adopt a mixing
angle in the range of $\phi_{2P}=-(3\sim 26)^\circ$, the decay width
of the $D_J(3000)$ can be explained. The mixing angle for
$D_J(3000)$ has an overlap with that for the $D_{sJ}(3040)$ in the
range of $\phi_{2P}=-(20\sim26)^\circ$. It is found that the mixing
angle $\phi_{2P}=-(20\sim26)^\circ$ obtained in present work is
close to the mixing angle $\phi_{2P}=-30.4^\circ$ suggested
in~\cite{Godfrey:2013aaa}.

If both $D_{sJ}(3040)$ and $D_J(3000)$ can be assigned as the mixed
state $|2P_1\rangle_L$, they should share nearly the same mixing
angle. Taking the same mixing angle we predict that the width of the
$D_J(3000)$ should be a little larger than that of the
$D_{sJ}(3040)$ for the larger decay phase space and more decay
channels of the $D_J(3000)$.

Recently, the strong decay properties of the $D_J(3000)$ were
studied in Ref.~\cite{Sun:2013qca} as well. Considering the
$D_J(3000)$ as the $D(|2P_1\rangle_L)$ with the mixing angle
$\phi_{2P}=-54.7^\circ$, they could well explain the measured width
of the $D_J(3000)$. However, it should be pointed out that they did
not include the $D(2600)\pi$ decay mode in their calculations.
Excluding the $D(2600)\pi$, our prediction is in agreement with that
in Ref.~\cite{Sun:2013qca}. To understand the nature of the
$D_J(3000)$, more observations are needed in future experiments.

\begin{widetext}
\begin{center}
\begin{figure}[ht]
\centering \epsfxsize=6.5 cm \epsfbox{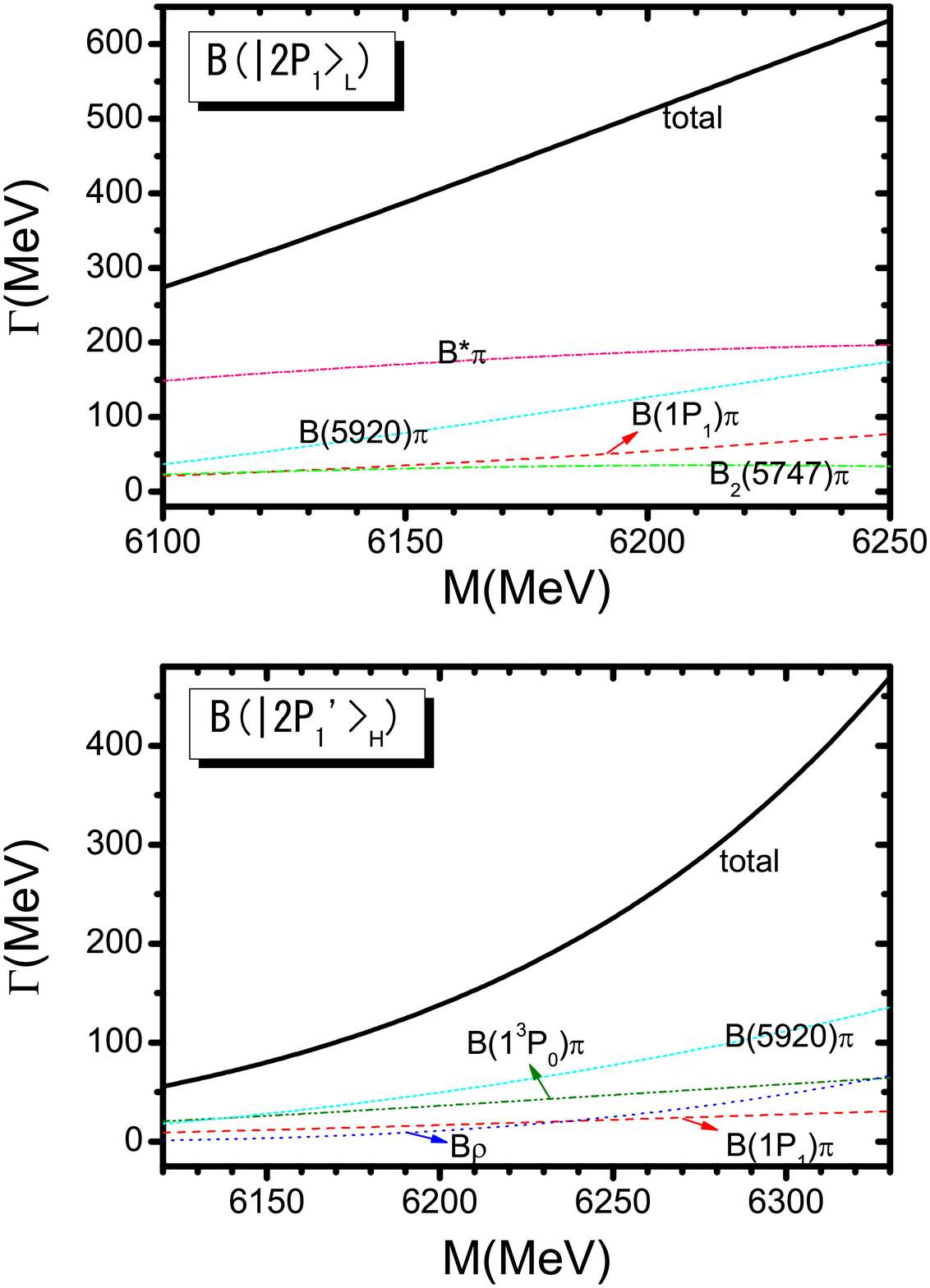} \epsfxsize=6.5 cm
\epsfbox{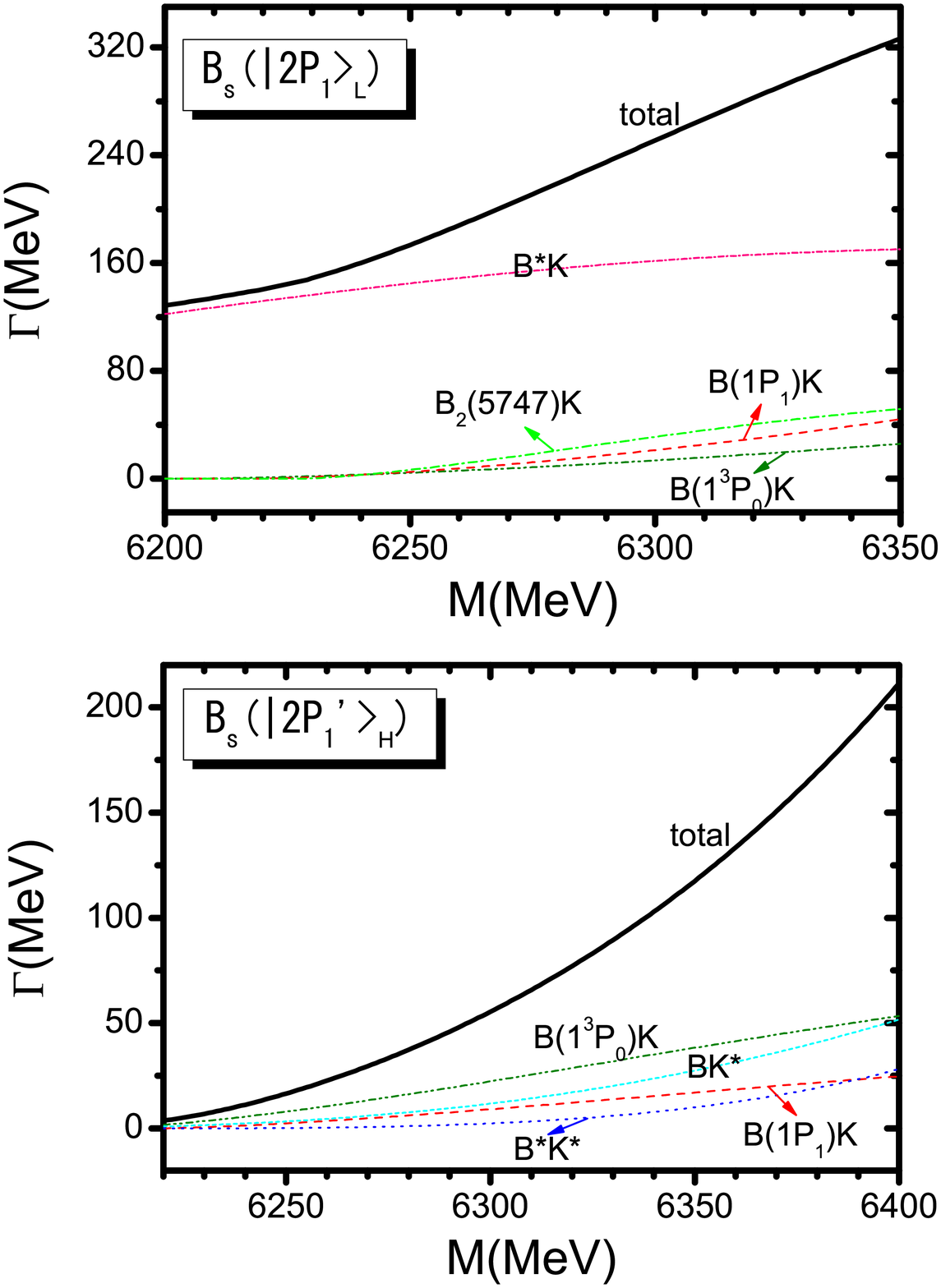}\caption{The decay width of the mixed states via
the $2^1P_1$-$2^3P_1$ mixing in the $B$- and $B_s$-meson families as
a function of the mass. Here, the mixing angle
$\phi_{2P}=-54.7^\circ$ is adopted. The $B(5920)$ stands for the
mixed state $B(|(SD)_1\rangle_L)\pi$. In the figure, we have hidden
some decay channels because of their small partial decay
widths.}\label{fig-PPBLM}
\end{figure}
\end{center}
\end{widetext}


In the $B$-meson family, the predicted mass for the low-mass state
$B(|2P_1\rangle_L)$ is $M\simeq 6.2$ GeV~\cite{Godfrey:1985xj,Di
Pierro:2001uu,Ebert:2009ua}, the mass of the high-mass state
$B(|2P_1'\rangle_H)$ is about $(20\sim 80)$ MeV heavier than that of
the low-mass state. Considering the uncertainty of the predicted
masses for these states, in Fig.~\ref{fig-PPBLM} we plot the partial
decay widths and total decay width of $B(|2P_1\rangle_L)$ and
$B(|2P_1'\rangle_H)$ as functions of the mass in the possible range,
where we adopt the mixing angle $\phi_{2P}=-54.7^\circ$ derived in
the heavy quark symmetry limit. From the figure, we find that the
low-mass state $B(|2P_1\rangle_L)$ is a broad state with a width of
$\Gamma\simeq (450\pm 150)$ MeV. The $B^*\pi$ decay mode governs its
strong decays, while the other decay channels
$B(|(SD)_1\rangle_L)\pi$, $B(1P_1)\pi$, $B(1^3P_2)\pi$, $B_s^*K$,
$B(1^3P_0)\pi$ and $B^*\eta$ also have sizable partial decay widths.

The width of the high-mass state $B(|2P_1' \rangle_H)$,
$\Gamma\simeq (50\sim450)$ MeV, is much narrower than that of the
low-mass state $B(|2P_1\rangle_L)$. The $B(|2P_1'\rangle_H)$
dominantly decays into $B(|(SD)_1\rangle_L)\pi$, $B(1^3P_0)\pi$,
$B(1P_1)\pi$ and $B(1^3P_2)\pi$. It might be a challenge to observe
the $B(|2P_1'\rangle_H)$ in these final states with higher
excitations of the $B$ meson.

Since the $D_{sJ}(3040)$ as a broad state has been observed in
experiments, the broad resonance $B(|2P_1\rangle_L)$ might be
observed in the $B^*\pi$ as well. To further know about the effects
of the uncertainties of the mixing angle on the strong decays of
$B(|2P_1\rangle_L)$, we have plotted the partial widths and total
decay width of $B(|2P_1\rangle_L)$ as functions of the mixing angle
$\phi_{2P}$ in Fig.~\ref{fig-PPBL1}, where we have fixed the mass of
$B(|2P_1\rangle_L)$ with $M=6175$ MeV predicted in~\cite{Di
Pierro:2001uu}. From the figure, it is seen that even the mixing
angle $\phi_{2P}$ varies in a large range
$\phi_{2P}\simeq-20^\circ\sim -80^\circ$, the strong decays of
$B(|2P_1\rangle_L)$ are still dominated by the $B^*\pi$ channel, and
the uncertainty of the decay width is no more than 60 MeV.


In the $B_s$-meson family, the predicted mass for the low-mass state
$B_s(|2P_1\rangle_L)$ is $M\simeq (6.28\sim 6.32)$
GeV~\cite{Godfrey:1985xj,Di Pierro:2001uu,Ebert:2009ua}, the mass of
the high-mass state $B_s(|2P_1'\rangle_H)$ is about $(20\sim 30)$
MeV heavier than that of the low-mass state.

\begin{widetext}
\begin{center}
\begin{figure}[ht]
\centering \epsfxsize=6.8 cm \epsfbox{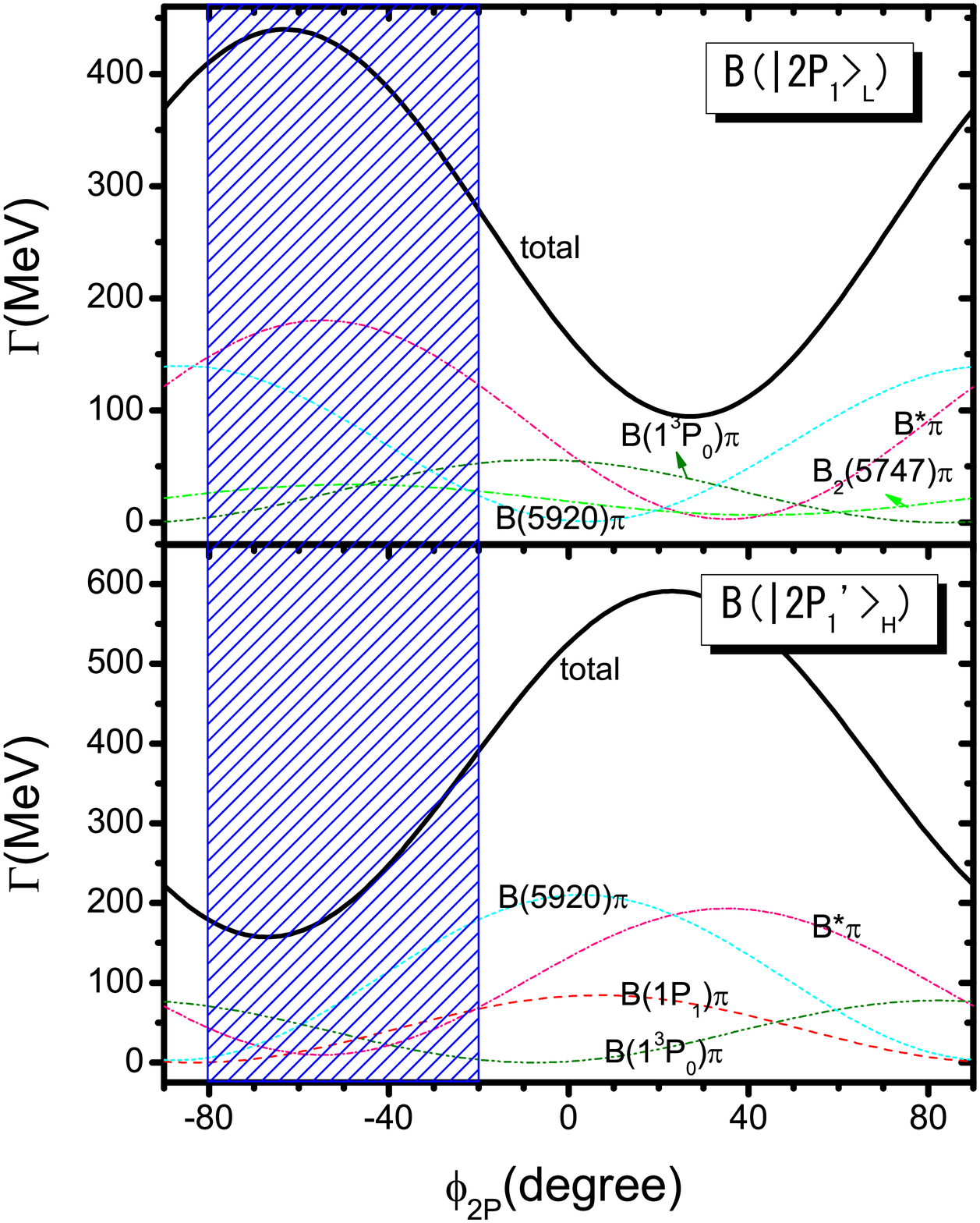} \epsfxsize=6.8 cm
\epsfbox{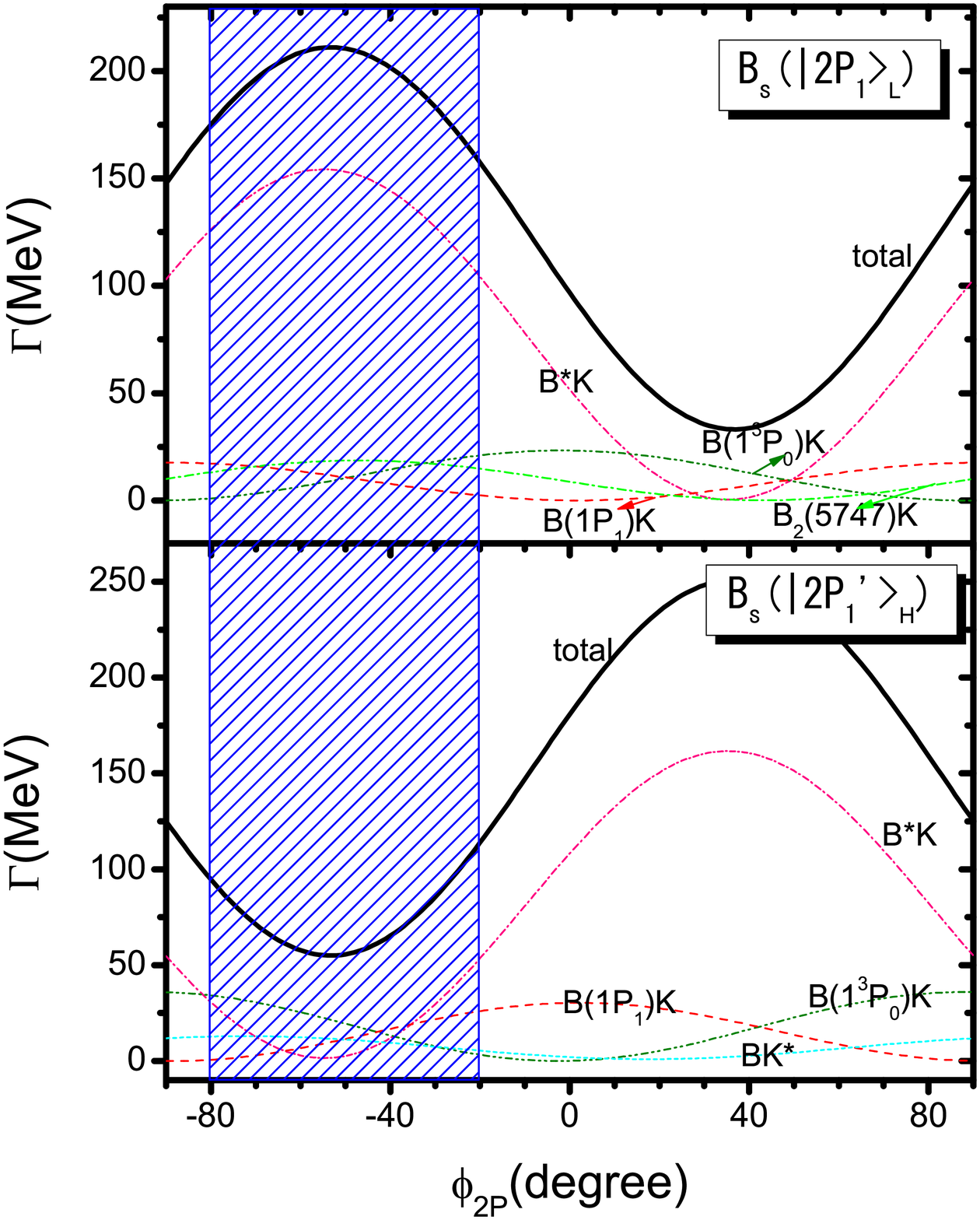} \caption{The partial decay width and total decay
width for the mixed states via $2^1P_1$-$2^3P_1$ mixing in the $B$-
and $B_s$-meson families as functions of the mixing angle
$\phi_{2P}$. The shaded region corresponds to the possible mixing
angle region derived from the strong decay properties of the
$D_{sJ}(3040)$ and $D_J(3000)$. The masses of the low-mass mixed
states $|2P_1\rangle_L$ in the $B$- and $B_s$-meson families are
adopted as 6.175 GeV and 6.275 GeV, respectively.  The masses of the
high-mass mixed states $|2P_1'\rangle_H$ in the $B$- and $B_s$-meson
families are adopted as 6.225 GeV and 6.30 GeV, respectively. In the
figure, we have hidden some decay channels because of their small
partial decay widths.}\label{fig-PPBL1}
\end{figure}
\end{center}
\end{widetext}

We have plotted the partial decay widths and total decay width of
$B_s(|2P_1\rangle_L)$ and $B_s(|2P_1'\rangle_H)$ as functions of the
mass with the mixing angle $\phi_{2P}=-54.7^\circ$ in
Fig.~\ref{fig-PPBLM}. From the figure, we find that the low-mass
state $B_s(|2P_1\rangle_L)$ has a broad width of $\Gamma\simeq
(220\pm 70)$ MeV within the predicted mass region $M\simeq (6.28\sim
6.32)$ GeV. The $B^*K$ decay channel governs its strong decays,
while the other decay channels $B(1P_1)K$, $B(1^3P_2)K$,
$B(1^3P_0)K$ also have sizable partial decay widths. The width of
the high-mass state $B_s(|2P_1'\rangle_H)$ is $\Gamma\simeq (95\pm
33)$ MeV within the possible mass region $M\simeq (6.30\sim 6.35)$
GeV, which is much narrower than that of $B_s(|2P_1\rangle_L)$. The
$B_s(|2P_1'\rangle_H)$ dominantly decays into the $B(1^3P_0)K$,
$B(1P_1)K$, and $BK^*$ channels. Although the $B_s(|2P_1'\rangle_H)$
has a relatively narrow width, it might be a great challenge to
observe this resonance in these final states with higher $B_s$
resonances.

The decay width of $B_s(|2P_1\rangle_L)$ is comparable with that of
the $D_{sJ}(3040)$. Thus, the $B_s(|2P_1\rangle_L)$ is most likely
to be observed in the $B^*K$ channel. To further know about the
effects of the uncertainties of the mixing angle on the strong
decays of $B_s(|2P_1\rangle_L)$, we have plotted the partial widths
and total decay width of $B_s(|2P_1\rangle_L)$ as functions of the
mixing angle $\phi_{2P}$ in Fig.~\ref{fig-PPBL1}, where we have
fixed the mass of $B_s(|2P_1\rangle_L)$ with $M=6275$ MeV. From the
figure, it is seen that even the mixing angle $\phi_{2P}$ is changed
in a large range $\phi_{2P}\simeq-20^\circ\sim -80^\circ$, the
strong decays of $B_s(|2P_1\rangle_L)$ are still dominated by the
$B^*K$ channel, and the uncertainty of the decay width is no more
than 50 MeV.

In summary, the strong decay properties of both $D_{sJ}(3040)$ and
$D_J(3000)$ can be explained by assigning them as the low-mass mixed
state $|2P_1\rangle_L$ via the $2^1P_1$-$2^3P_1$ with a mixing angle
in the range of $\phi_{2P}=-(20\sim26)^\circ$. The mixed state
$B_s(|2P_1\rangle_L)$ is a narrow state, which is most likely to be
observed in the $B^*K$ channel. The mixed state $B(|2P_1\rangle_L)$
might have a broad width; thus, its discovery potentials might be
small. The high-mass mixed states $D(|2P_1'\rangle_H)$,
$D_s(|2P_1'\rangle_H)$, $B(|2P_1'\rangle_H)$, and
$B_s(|2P_1'\rangle_H)$ are usually narrower than those of low-mass
states. These states dominantly decay into the first $P$-wave states
by emitting a light pseudoscalar meson. Observations in the final
states containing a low-lying $P$-wave state are expected to be
carried out in future experiments.

\begin{widetext}
\begin{center}
\begin{figure}[ht]
\centering \epsfxsize=15.0 cm \epsfbox{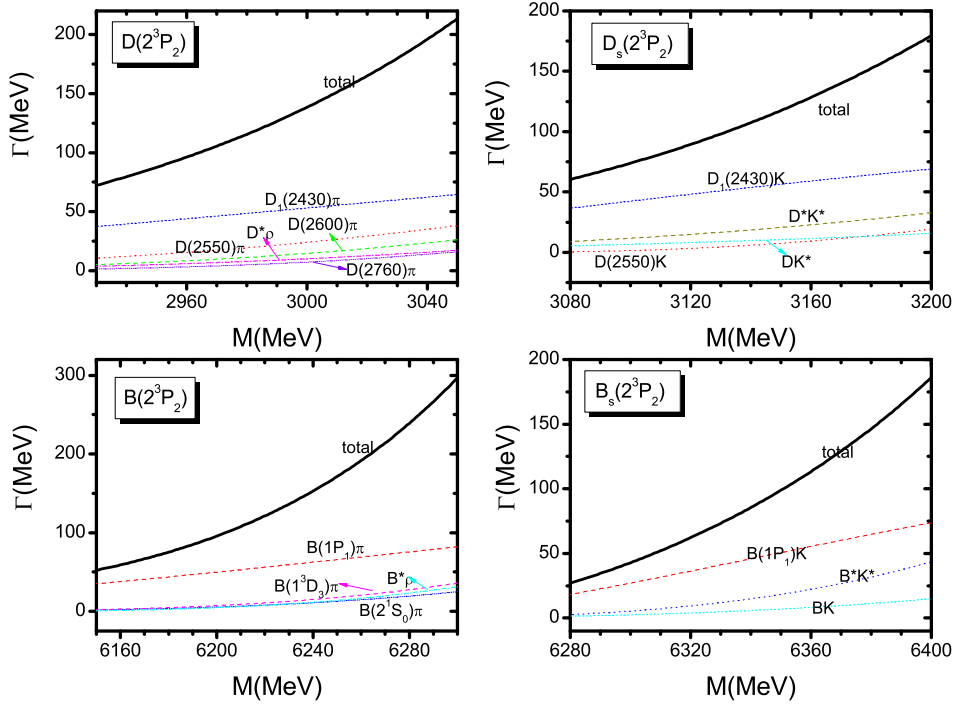} \caption{The partial
decay width and total decay widths of the $2^3P_{2}$ states in the
heavy-light mesons as functions of the mass. In the figure, we have
hidden some decay channels because of their small partial decay
widths. The mass of the $B(1^3D_3)$ is adopted as $M=5970$
MeV.}\label{fig-3P2}
\end{figure}
\end{center}
\end{widetext}

\subsection{The $2^3P_2$ states}


In the $D$-meson family, the predicted mass of the $2^3P_2$ is about
$3.02$ GeV~\cite{Di Pierro:2001uu, Ebert:2009ua,Sun:2014wea}. To
know about the decay properties of the $D(2^3P_2)$, in
Fig.~\ref{fig-3P2} we plot its decay width as a function of mass in
the range of $M=(2.95\sim3.05)$ GeV. From the figure, it is found
that the $D(2^3P_2)$ mainly decays into the $D_1(2430)\pi$,
$D(2550)\pi$ and $D(2600)\pi$ channels. The predicted total decay
width of $D(2^3P_2)$ is $\Gamma\simeq 150$ MeV, if we adopt the mass
$\sim 3.01$ GeV as predicted in Ref.~\cite{Ebert:2009ua}.

Considering the newly observed state $D_J^*(3000)$ as a candidate of
$D(2^3P_2)$, we find that the observed mass, width ($\Gamma\simeq
110$ MeV) and natural parity of the
$D_J^*(3000)$~\cite{Aaij:2013sza} can be well explained. However,
the $D\pi$ decay channel is not the main decay channel of
$D(2^3P_2)$, which is tiny compared with the dominant decay mode
$D_1(2430)\pi$. The predicted branching fraction is
\begin{eqnarray}
\frac{\Gamma[D\pi]}{\Gamma_{\mathbf{total}}}\simeq 1\sim 2\%,
\end{eqnarray}
which is compatible with the predictions in
Refs.~\cite{Sun:2013qca,Lu:2014zua}. To clarify the nature of the
$D_J^*(3000)$, it is suggested to further observe this state in the
$D_1(2430)\pi$ channel. If the $D_J^*(3000)$ corresponds to the
$D(2^3P_2)$ state, it should be found in the $D_1(2430)\pi$ channel
as well.


In the $D_s$-meson family, the predicted mass of the $2^3P_2$ is
about $3.15$ GeV~\cite{Di Pierro:2001uu, Ebert:2009ua,Sun:2014wea}.
To know about the decay properties of the $D_s(2^3P_2)$, in
Fig.~\ref{fig-3P2} we plot its decay width as a function of mass in
the range of $M=(3.1\sim3.2)$ GeV. From the figure, it is found that
the $D_s(2^3P_2)$ mainly decays into the $D_1(2430)K$ channel. The
total decay width of $D_s(2^3P_2)$ is $\Gamma\simeq 120\pm 60$ MeV.
This resonance might be observed in the $D_1(2430)K$ final states.


In the $B$-meson family, the predicted mass of the $2^3P_2$ is about
$6.19\sim 6.26$ GeV~\cite{Di Pierro:2001uu, Ebert:2009ua}. In
Fig.~\ref{fig-3P2}, we have plotted its decay width as a function of
mass in the range of $M=(6.15\sim 6.3)$ GeV. From the figure it is
found that the total decay width of $B(2^3P_2)$ is $\Gamma\simeq
(50\sim300)$ MeV. The $B(2^3P_2)$ mainly decays into the
$B_1(5721)\pi$ channel.


In the $B_s$-meson family, the predicted mass of the $2^3P_2$ is
about $6.29\sim 6.36$ GeV~\cite{Di Pierro:2001uu,
Ebert:2009ua,Sun:2014wea}. The decay properties of the $B(2^3P_2)$
have been shown in Fig.~\ref{fig-3P2}. From the figure, it is found
that the $B_s(2^3P_2)$ mainly decays into $B_1(5721)K$. The total
decay width of $B_s(2^3P_2)$ is $\Gamma\simeq (30\sim 180)$ MeV.
This state might be observed in the $B_1(5721)K$ channel.

In a brief, the decay widths of the $2^3P_2$ excitations in the
heavy-light spectroscopy are not very broad, they have good
potentials to be observed in their dominant decay channels.

\begin{widetext}
\begin{center}
\begin{figure}[ht]
\centering \epsfxsize=15.0 cm \epsfbox{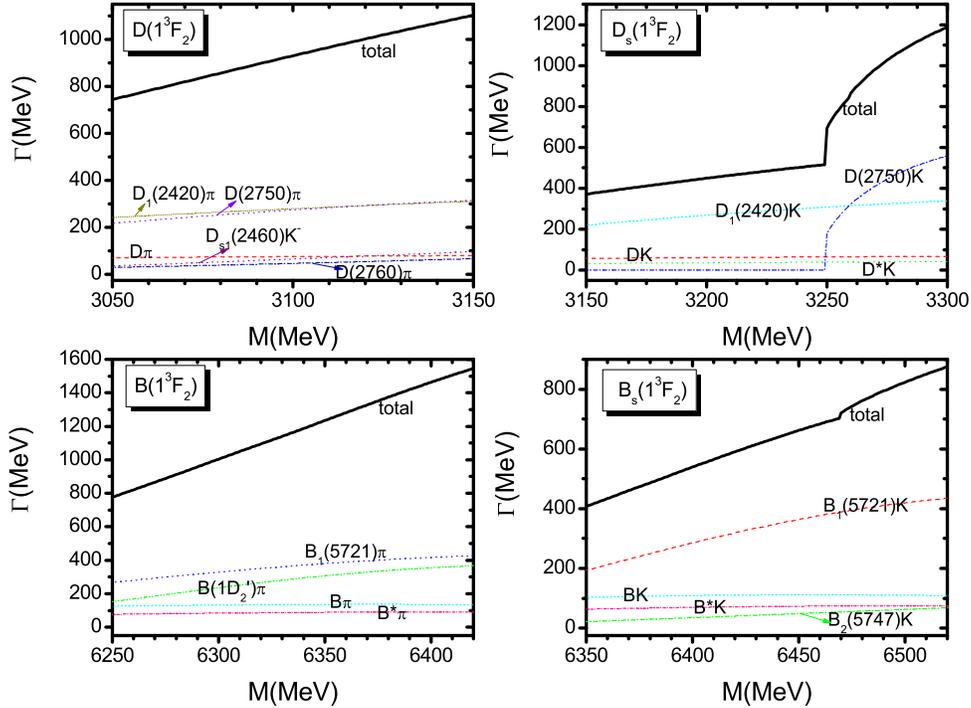} \caption{The partial
decay width and total decay widths of the $1^3F_{2}$ states in the
heavy-light mesons as functions of the mass. In the figure, we have
hidden some decay channels because of their small partial decay
widths. The mass of the $B(1D_2')$ is adopted as $M=6025$
MeV.}\label{fig-3F2}
\end{figure}
\end{center}
\end{widetext}

\subsection{The $1^3F_2$ states}


In the $D$-meson family, the predicted mass for the $1^3F_2$ state
is $\sim3.1$ GeV~\cite{Di Pierro:2001uu, Ebert:2009ua,Sun:2014wea}.
Considering the uncertainties of the mass, in Fig.~\ref{fig-3F2} we
plot the partial decay widths and total width as functions of mass
in the range of $M=(3.05\sim 3.15)$ GeV. It is found that the
$1^3F_2$ state in the $D$-meson family is a very broad state with a
width of $\Gamma\simeq 900\pm 200$ MeV. This state dominantly decays
into the $D(2420)\pi$, $D(2750)\pi$ and $D\pi$ channels. The partial
widths for the $D_s(2460)\pi$, $D^*\pi$ and $D(2760)\pi$ are also
sizable.


In the $D_s$-meson family, the predicted mass for the $1^3F_2$ state
is $\sim3.23$ GeV~\cite{Di Pierro:2001uu, Ebert:2009ua,Sun:2014wea}.
Considering the uncertainties of the mass, we vary the mass in the
range of $M=(3.15\sim 3.30)$ GeV, the strong decay properties of the
$1^3F_2$ state have been shown in Fig.~\ref{fig-3F2} as well. From
the figure it is seen that the decay width of the $D_s(1^3F_2)$
state is $\Gamma\sim 500$ MeV. The main decay channels are
$D(2420)K$, $DK$ and $D^*K$. If the $D(2750)K$ channel is opened,
the $D(2750)K$ mode will dominate the decays, the decay width of the
$D_s(1^3F_2)$ state will become very broad with a width of
$\Gamma>700$ MeV.


In the $B$-meson family, the predicted mass for the $1^3F_2$ state
is $6.26\sim6.41$ GeV~\cite{Di Pierro:2001uu,
Ebert:2009ua,Sun:2014wea}. In this mass region, we show the strong
decay properties of the $B(1^3F_2)$ state in Fig.~\ref{fig-3F2}. It
is found that this state is also a very broad state with a width of
$\Gamma\simeq 1200\pm 400$ MeV. It mainly decays into the
$B(1P_1)\pi$, $B(1D_2')\pi$, $B\pi$ and $B^*\pi$ channels.



In the $B_s$-meson family, the predicted mass for the $1^3F_2$ state
is $6.37\sim6.50$ GeV~\cite{Di Pierro:2001uu, Ebert:2009ua}. In this
mass region, we study the strong decays of the $B_s(1^3F_2)$ state,
our results are shown in Fig.~\ref{fig-3F2} as well. It is seen that
this state is a very broad state with a width of $\Gamma\simeq
400\sim 900$ MeV. The strong decays are governed by the $B(1P_1)K$.
The other main decay channels are $BK$, $B^*K$ and $B(1^3P_2)K$.

As a whole, the $1^3F_2$ excitations in the heavy-light mesons are
very broad resonances, whose width is larger than 400 MeV. Such
broad states might be difficult to observe in experiments, which
might explain why these states are still missing in the heavy-light
meson spectroscopy.

\begin{widetext}
\begin{center}
\begin{figure}[ht]
\epsfxsize=7 cm \epsfbox{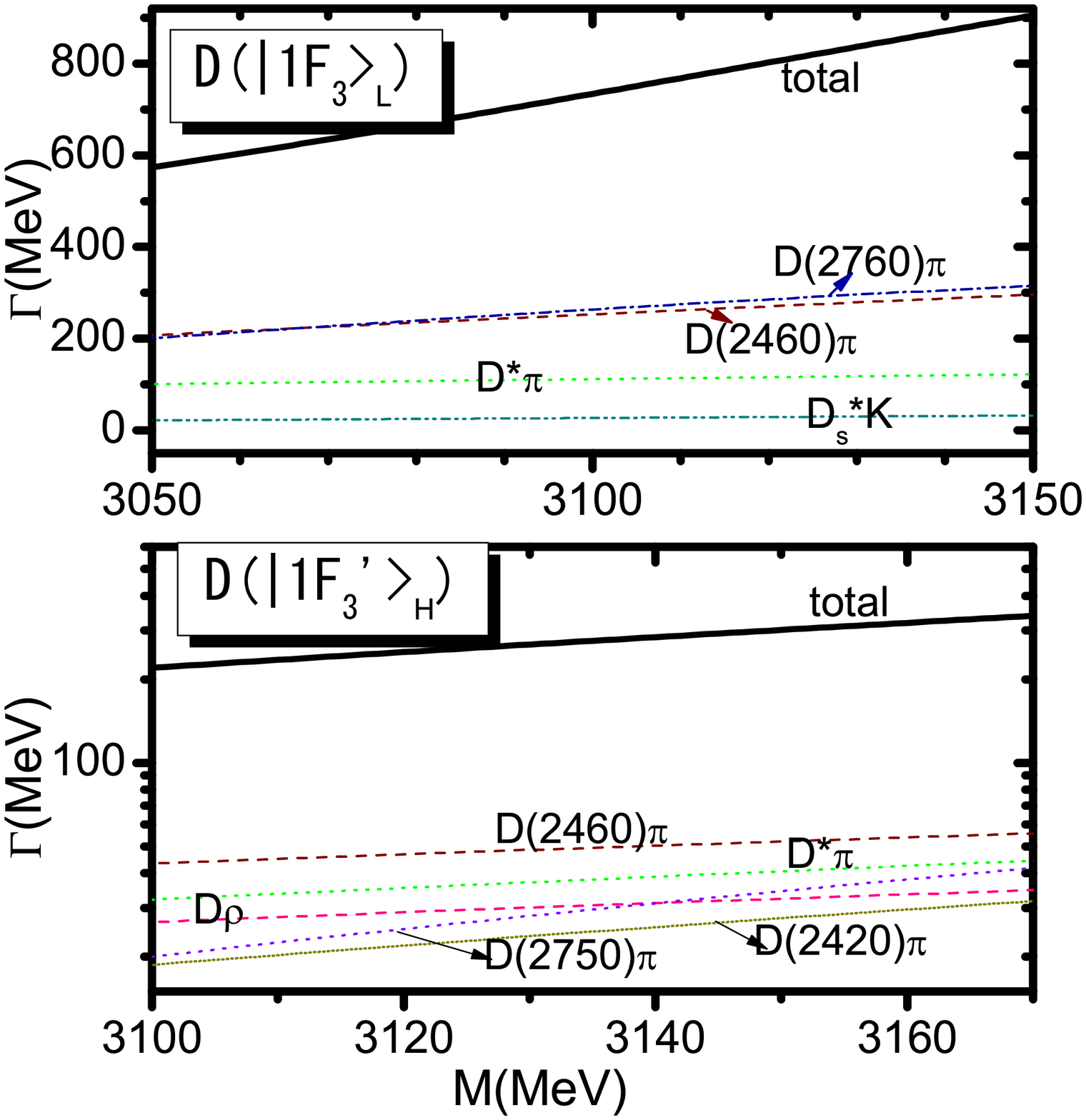} \epsfxsize=7 cm
\epsfbox{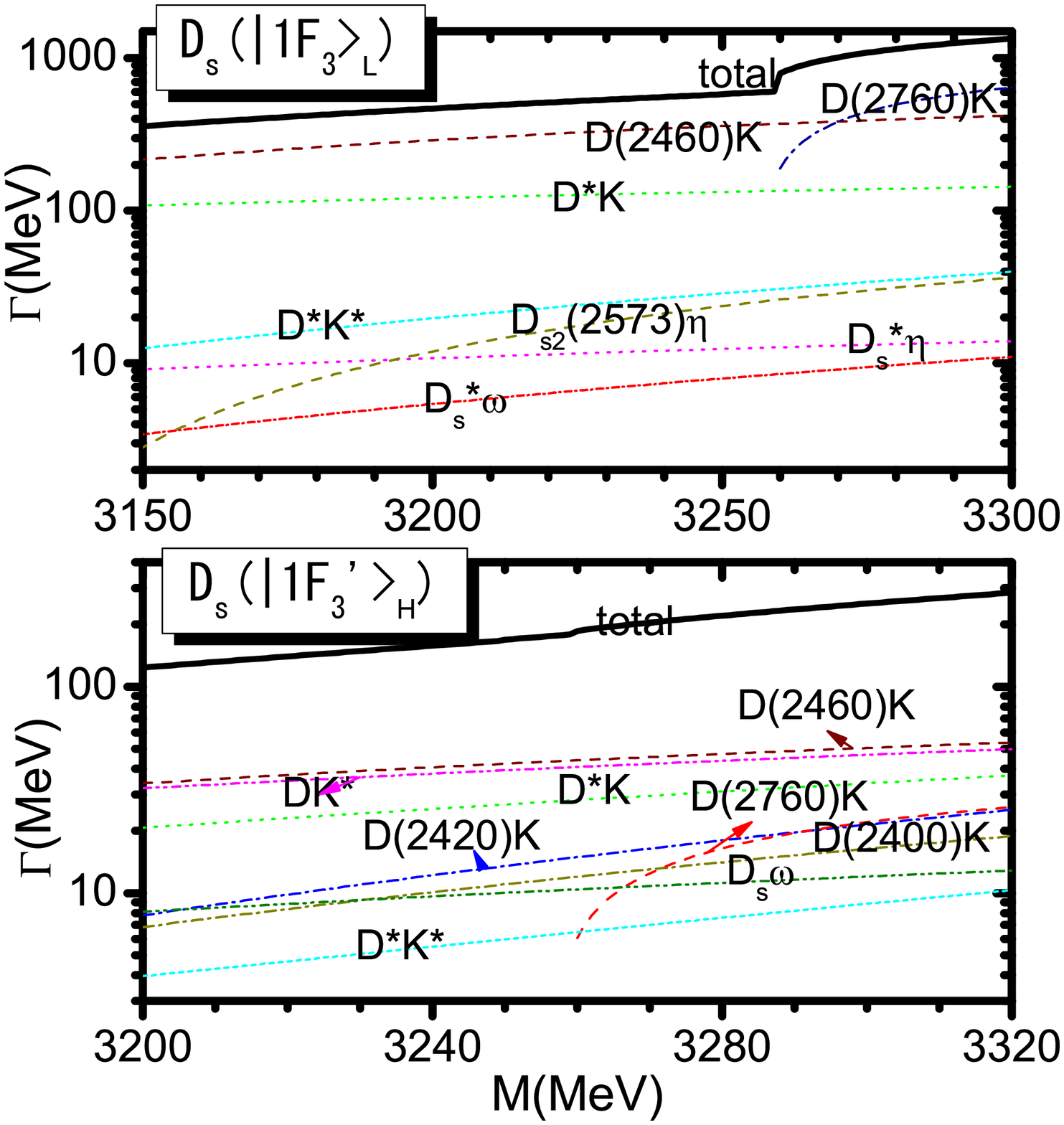} \epsfxsize=7 cm \epsfbox{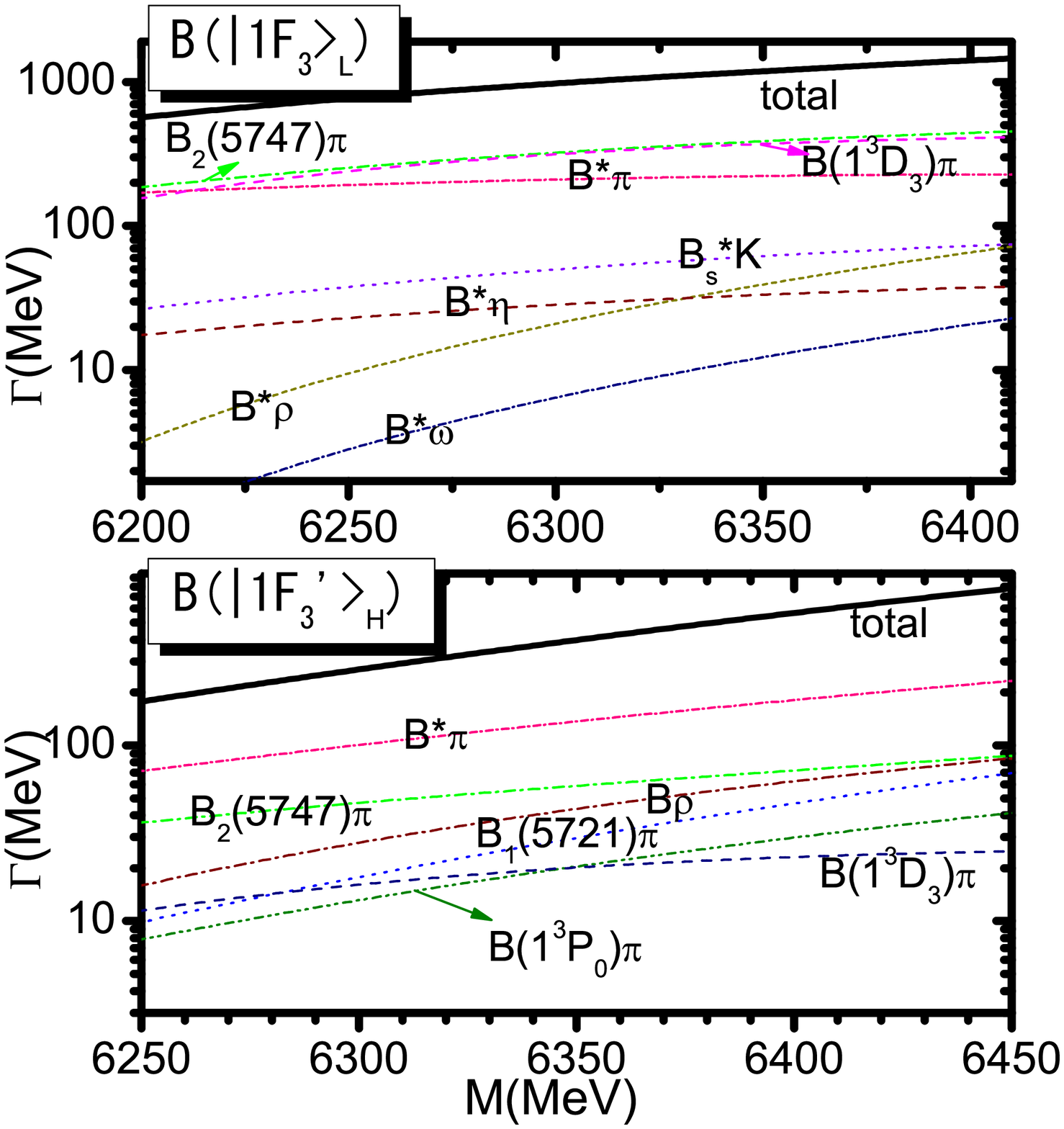} \epsfxsize=7
cm \epsfbox{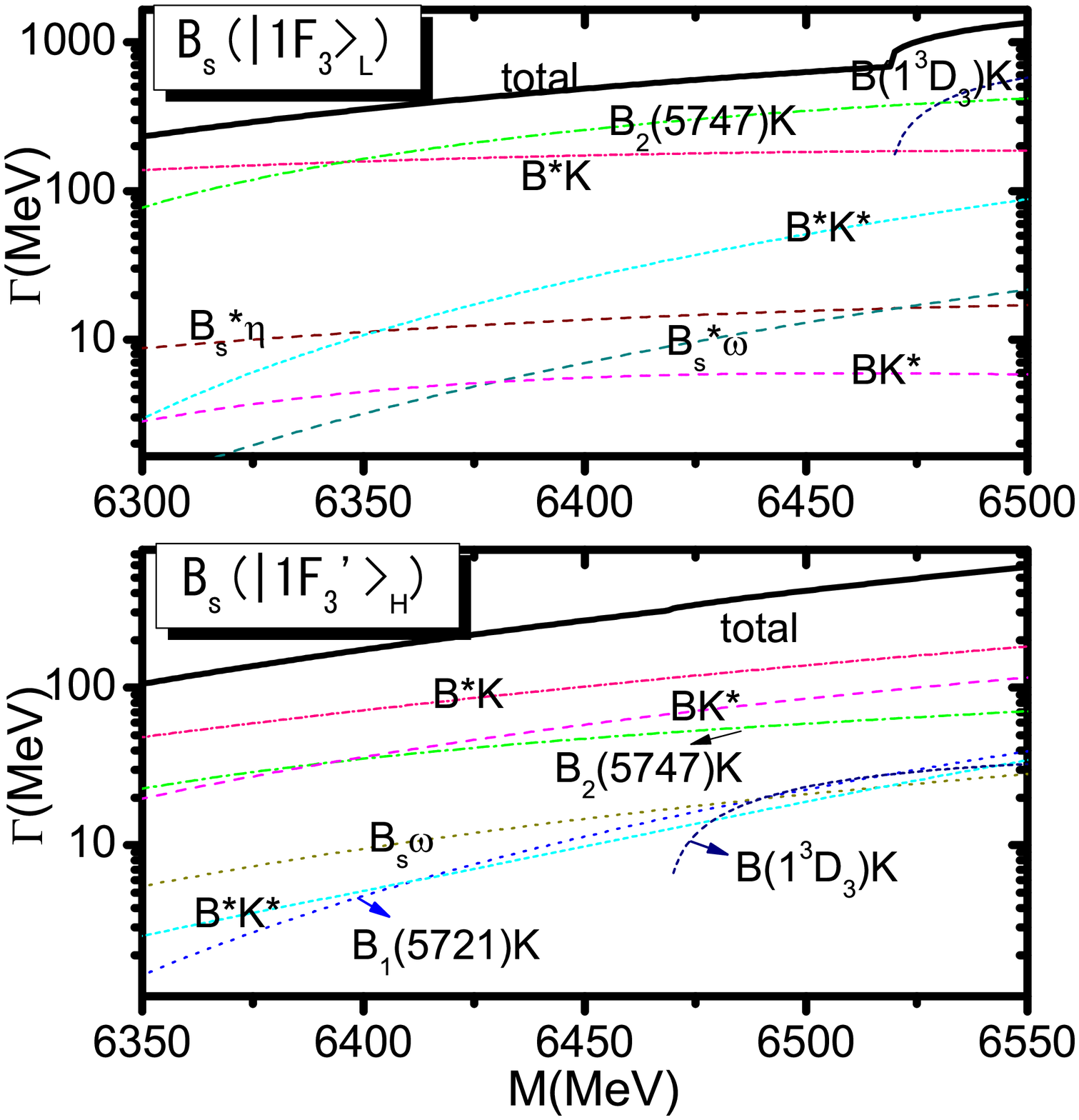}\caption{ The partial decay width and total
decay width for the mixed states via the $1^1F_3$-$1^3F_3$ mixing as
functions of the mass. The mixing angle is adopted as
$\phi_{1F}=-49.1^\circ$. In the figure, we have hidden some decay
channels because of their small partial decay
widths.}\label{fig-3F3M}
\end{figure}
\end{center}
\end{widetext}

\subsection{The $1^3F_3$-$1^3F_3$ mixing}

Since the heavy-light mesons are not charge conjugation eigenstates,
state mixing between spin $\mathbf{S}=0$ and $\mathbf{S}=1$ states
with the same $J^P$ can occur via  the spin-orbit interactions
\cite{Godfrey:1986wj,Close:2005se,Swanson:2006st}. The physical
states with $J^P=3^+$ can then be described as
\begin{equation}\label{mixs}
\left(\begin{array}{c}|1F_{3}\rangle_L\cr |1F_{3}'\rangle_H
\end{array}\right)=\left(\begin{array}{cc} \cos\phi_{1F} &\sin\phi_{1F}\cr -\sin\phi_{1F} & \cos\phi_{1F}
\end{array}\right)
\left(\begin{array}{c} |1^1F_3\rangle \cr |1^3F_3\rangle
\end{array}\right),
\end{equation}
where the subscriptions $L$ and $H$ stand for the low mass and high
mass of the physical states after the mixing. In the heavy quark
symmetry limit, the mixing angle $\phi_{1F}$ is predicted to be
$\phi_{1F}\simeq -49.1^\circ$~\cite{Ebert:2009ua}. It should be
mentioned that with the mixing angle $\phi_{1F}\simeq -49.1^\circ$,
the low-mass state $|1F_{3}\rangle_L$ usually has a broad width,
while the high-mass state $|1F_{3}'\rangle_H$ has a narrower width.


In the $D$-meson family, the predicted masses for the mixed states
$|1F_{3}\rangle_L$ and $|1F_{3}\rangle_H$ are about $3.07\sim3.13$
GeV and $3.12\sim 3.15$ GeV, respectively~\cite{Di Pierro:2001uu,
Ebert:2009ua}. With the mixing angle $\phi_{1F}\simeq -49.1^\circ$
derived in the heavy quark symmetry limit, we predicted the strong
decay properties of the mixed states $D(|1F_{3}\rangle_L)$ and
$D(|1F_{3}'\rangle_H)$ in their possible mass region. The results
are shown in Fig.~\ref{fig-3F3M}. It is found that the
$D(|1F_{3}\rangle_L)$ is a very broad state with a width of
$\Gamma\simeq 600\sim 900$ MeV. Its strong decays are dominated by
the $D(2760)\pi$, $D(2460)\pi$ and $D^*\pi$ channels. The width of
the $D(|1F_{3}'\rangle_H)$ is $\Gamma\simeq 200\sim 300$ MeV, which
is much narrower than that of the low-mass state. The
$D(|1F_{3}'\rangle_H)$ mainly decays into the $D(2460)\pi$,
$D^*\pi$, $D\rho$, $D(2420)\pi$ and $D(2750)\pi$ channels.


In the $D_s$-meson family, the predicted masses for the mixed states
$|1F_{3}\rangle_L$ and $|1F_{3}'\rangle_H$ are about $3.20\sim3.26$
GeV and $3.25\sim 3.27$ GeV, respectively~\cite{Di Pierro:2001uu,
Ebert:2009ua}. The strong decay properties of these mixed states are
also studied in their possible mass region. Our predictions are
shown in Fig.~\ref{fig-3F3M}.  It is seen that the width of the
low-mass mixed state $D_s(|1F_{3}\rangle_L)$ is very broad, which is
in the range of $\Gamma\simeq 400\sim 800$ MeV. Its strong decays
are governed by the $D(2460)K$ and $D^*K$ channels. The high-mass
state $D_s(|1F_{3}'\rangle_H)$ has a much narrower decay width,
i.e., $\Gamma\simeq 100\sim250$ MeV. This high-mass state dominantly
decays into the $D(2460)K$, $D^*K$, and $DK^*$ channels.

\begin{widetext}
\begin{center}
\begin{figure}[ht]
\epsfxsize=7.0 cm \epsfbox{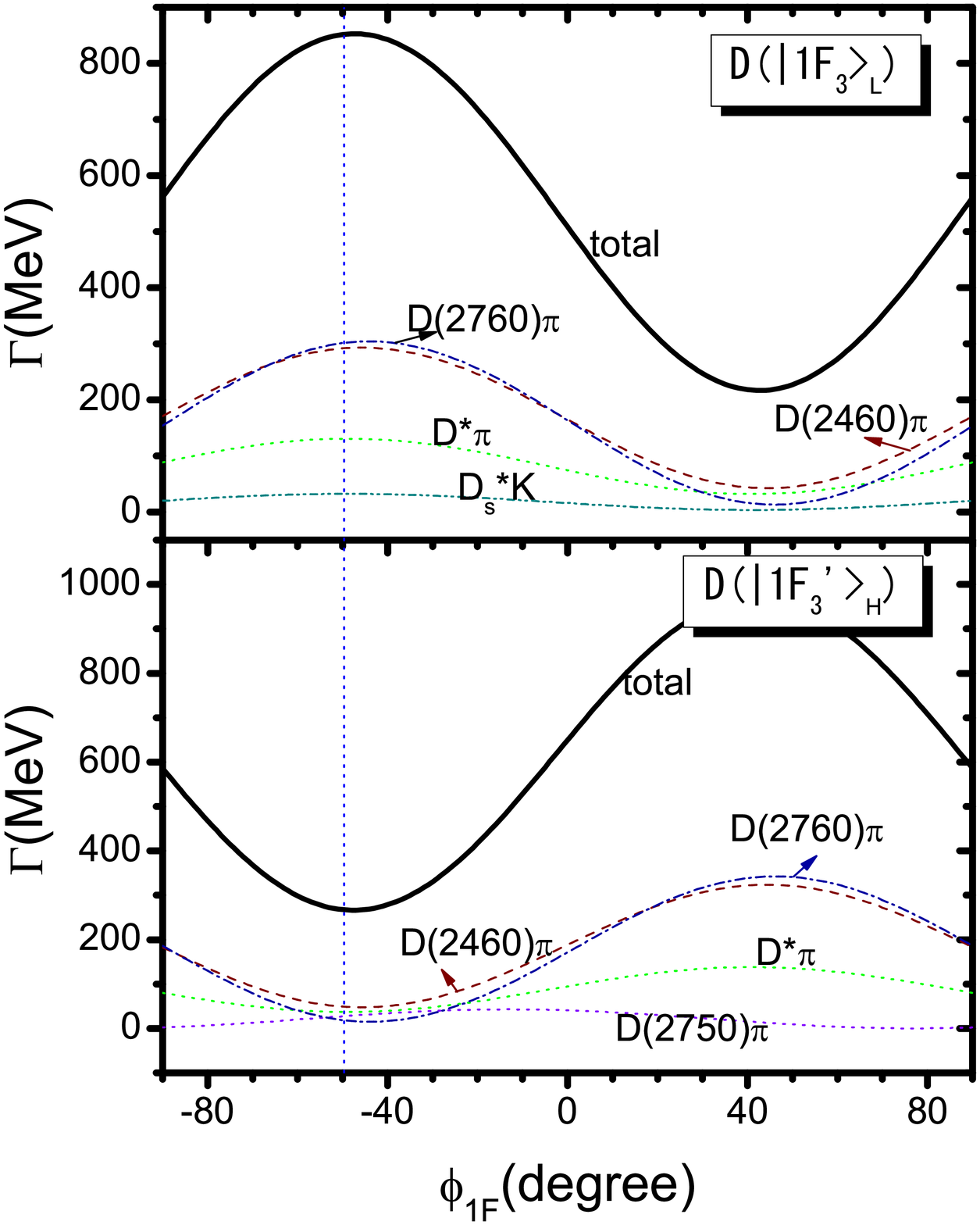} \epsfxsize=7.0 cm
\epsfbox{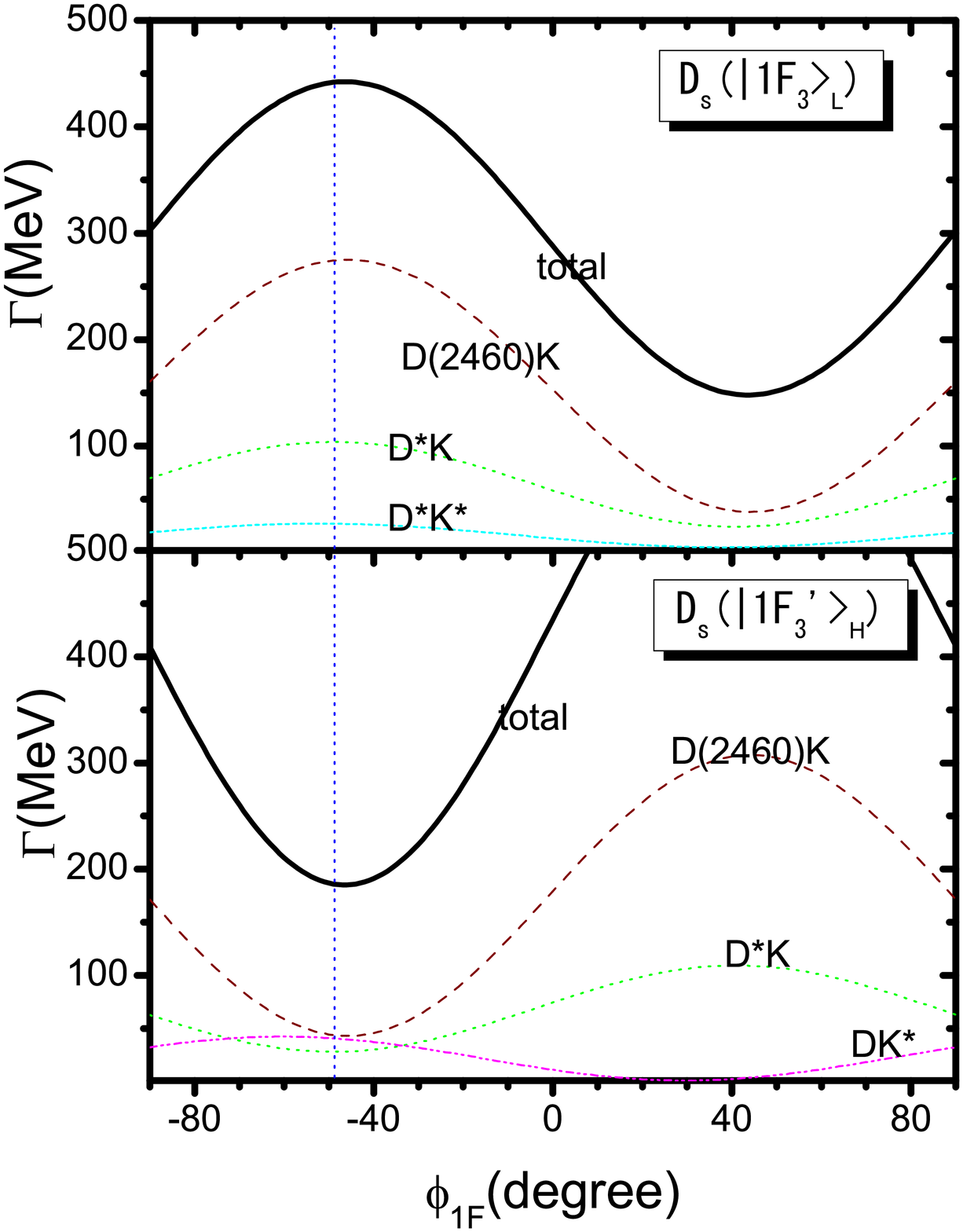} \epsfxsize=7.0 cm \epsfbox{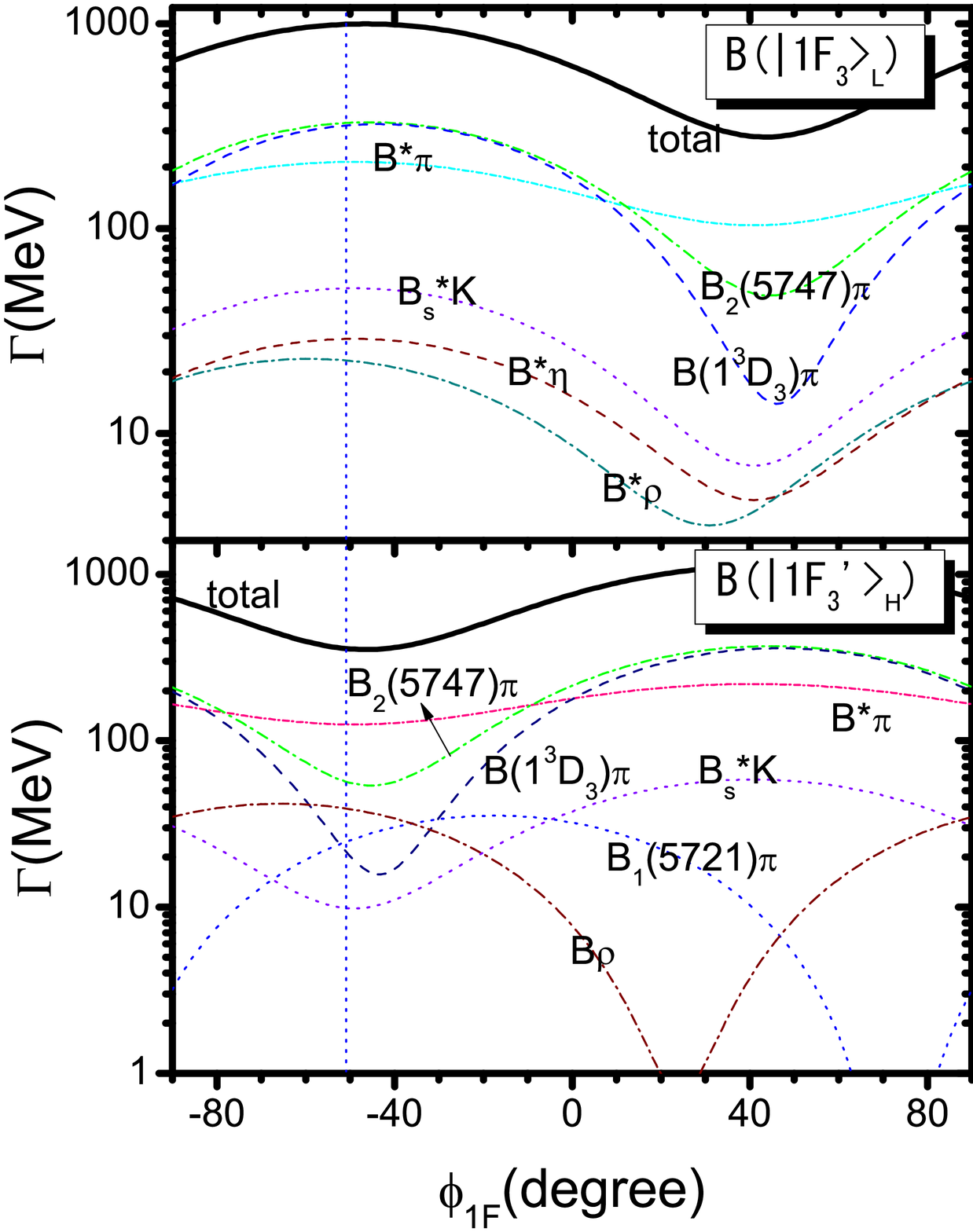}
\epsfxsize=7.0 cm \epsfbox{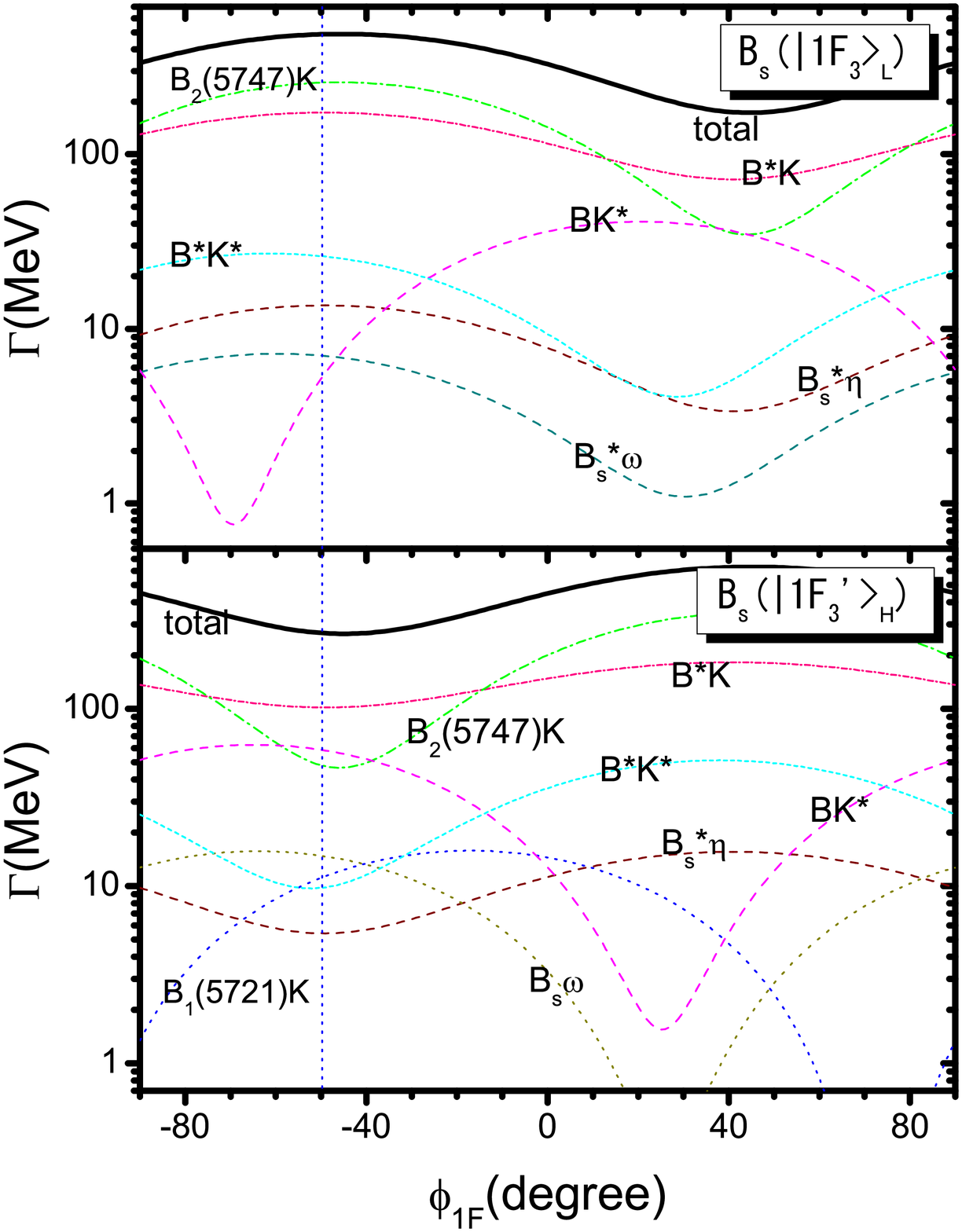} \caption{The partial decay
width and total decay width for the mixed states via the
$1^1F_3$-$1^3F_3$ mixing as functions of the mixing angle. The
masses for the low-mass mixed states $D(|1F_{3}\rangle_L)$,
$D_s(|1F_{3}\rangle_L)$, $B(|1F_{3}\rangle_L)$ and
$B_s(|1F_{3}\rangle_L)$ are adopted as 3100, 3230, 6305, and 6400
MeV, respectively. The masses for the high-mass mixed states
$D(|1F_{3}'\rangle_H)$, $D_s(|1F_{3}'\rangle_H)$,
$B(|1F_{3}'\rangle_H)$ and $B_s(|1F_{3}'\rangle_H)$ are adopted as
3130, 3260, 6335, and 6450 MeV, respectively. In the figure, we have
hidden some decay channels for their small partial decay
widths.}\label{fig-3F3}
\end{figure}
\end{center}
\end{widetext}


In the $B$-meson family, the predicted masses for the mixed states
$|1F_{3}\rangle_L$ and $|1F_{3}'\rangle_H$ are about $6.22\sim6.39$
GeV and $6.27\sim 6.42$ GeV, respectively~\cite{Di Pierro:2001uu,
Ebert:2009ua}. In the possible mass region, we study the strong
decay properties of these mixed states in the $B$-meson family. The
results are shown in Fig.~\ref{fig-3F3M}. It is found that the
low-mass mixed state $B(|1F_{3}\rangle_L)$ is a very broad state
with a width of $\Gamma\simeq 650\sim1400$ MeV. This state
dominantly decays into the $B_2(5747)\pi$, $B(1^3D_3)\pi$, and
$B^*\pi$ channels. While for the $B(|1F_{3}'\rangle_H)$, the decay
width is $\Gamma\simeq 200\sim650$ MeV, which is sensitive to the
mass. If the $B(|1F_{3}'\rangle_H)$ state has a smaller mass as
predicted in Ref.~\cite{Di Pierro:2001uu}, it might be observed in
its main decay channels $B^*\pi$ and $B_2(5747)\pi$.


In the $B_s$-meson family, the predicted masses for the mixed states
$|1F_{3}\rangle_L$ and $|1F_{3}'\rangle_H$ are about $6.33\sim6.47$
GeV and $6.38\sim 6.52$ GeV, respectively~\cite{Di Pierro:2001uu,
Ebert:2009ua}. Considering the uncertainties of the mass, we have
plotted the strong decay properties of these mixed states as
functions of the mass in Fig.~\ref{fig-3F3M}. From the figure, it is
seen that the $|1F_{3}\rangle_L$ state in the $B_s$-meson family has
a width of $\Gamma\simeq 200\sim 700$ MeV, whose strong decays are
dominated by the $B_2(5747)K$ and $B^*K$ channels. The high-mass
state $|1F_{3}\rangle_H$ has a relatively smaller width, which is
$\Gamma\simeq 100\sim 500$ MeV. Its strong decays are governed by
the $B^*K$, $BK^*$, and $B_2(5747)K$ channels. These mixed states
$|1F_{3}\rangle_L$ and $|1F_{3}'\rangle_H$ in the $B_s$-meson family
might be observed in the $B^*K$ channel, if they have a smaller mass
as predicted in Ref.~\cite{Di Pierro:2001uu}.

In the calculations we have adopted the ideal mixing angle
$\phi_{1F}\simeq -49.1^\circ$ extracted from the heavy quark
symmetry limit. This ideal mixing angle might have some
uncertainties. To see the effects of the uncertainties of the mixing
angle on the strong decay properties, we have plotted the partial
decay widths and total decay width as functions of the mixing angle
in Fig.~\ref{fig-3F3}. From the figure, it is seen that strong decay
properties of the mixed states do not change obviously, if we
consider an uncertainty $\pm 20^\circ$ around the ideal mixing angle
$\phi_{1F}\simeq -49.1^\circ$.

In summary, the mixed states $|1F_{3}\rangle_L$ in the heavy-light
mesons are usually broad states, they should be difficult to find in
experiments. The decay width of the high-mass mixed states
$|1F_{3}'\rangle_H$ is relatively narrower than that of the low-mass
states. These high-mass states $|1F_{3}'\rangle_H$ might be observed
in future experiments if they have a smaller mass as predicted in
Ref.~\cite{Di Pierro:2001uu}. However, these high-mass states
$|1F_{3}'\rangle_H$ might be very broad states if their masses are
as large as those predicted in Ref.~\cite{Ebert:2009ua}.

\begin{widetext}
\begin{center}
\begin{figure}[ht]
\centering \epsfxsize=16.0 cm \epsfbox{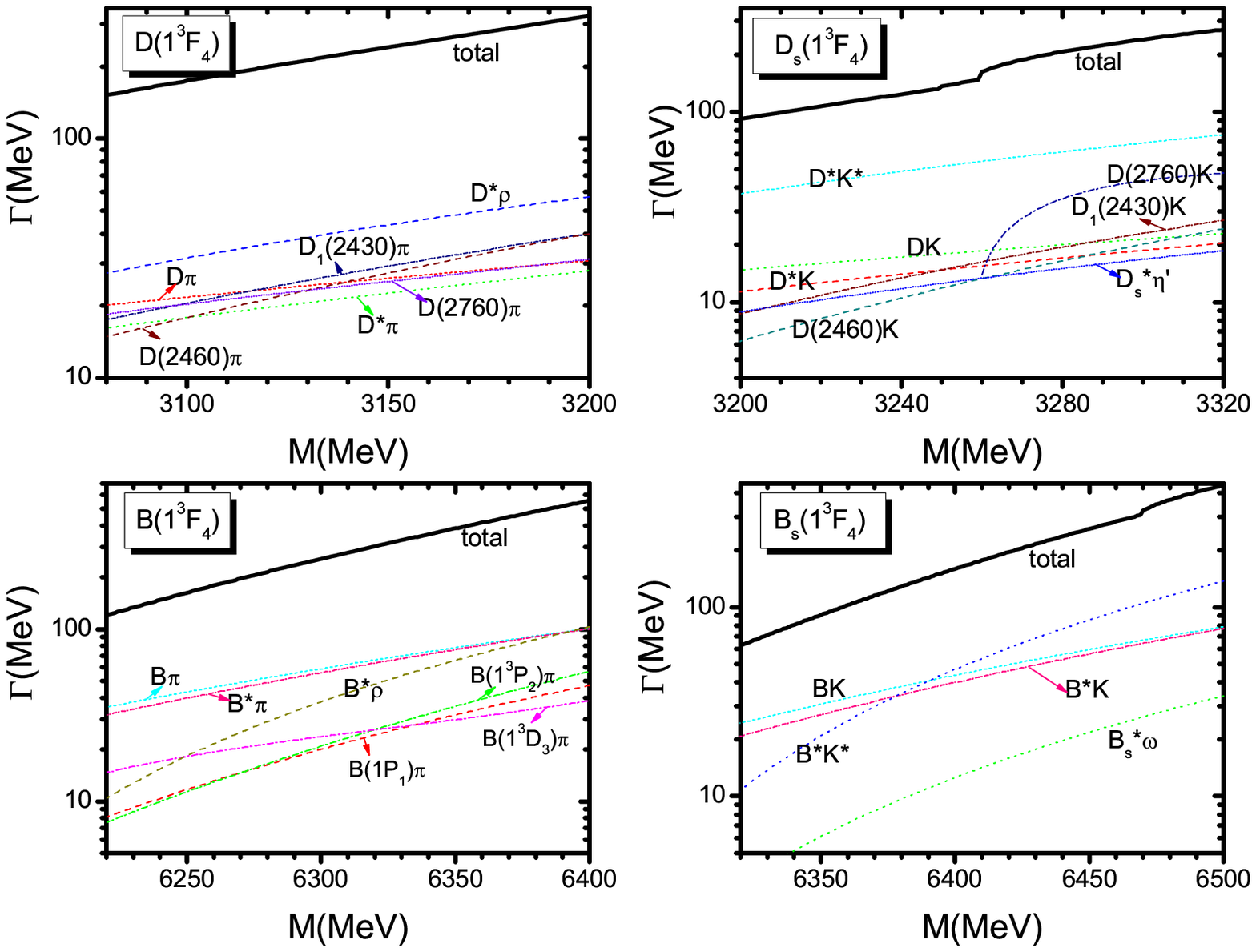} \caption{The partial
decay width and total decay width for the $1^3F_4$ excitations as
functions of the mass. In the figure, we have hidden some decay
channels because of their small partial decay
widths.}\label{fig-3F4}
\end{figure}
\end{center}
\end{widetext}

\subsection{The $1^3F_4$ states}


In the $D$-meson family, the predicted mass for the $1^3F_4$ state
is $3.09\sim3.19$ GeV~\cite{Di Pierro:2001uu,
Ebert:2009ua,Sun:2014wea}. In this mass region, the predicted strong
properties are shown in Fig.~\ref{fig-3F4}. The width of the
$D(1^3F_4)$ state is $\Gamma\simeq 230\pm 70$ MeV. It dominantly
decays into the $D^*\rho$, $D\pi$, $D(2430)\pi$, $D(2460)\pi$,
$D^*\pi$ and $D(2760)\pi$ channels. The branching ratios for
$D^*\pi$ and $D\pi$ are 10\% and 11\%, respectively.

We find that if taking the newly observed natural parity resonance
$D^*_J(3000)$ as a candidate of the $D(1^3F_4)$, the strong decay
properties of $D^*_J(3000)$ could be well explained (see
Tab.~\ref{Tab:D13F4}). However, the predicted mass is about $100\sim
200$ MeV larger than the data. If the $D^*_J(3000)$ corresponds to
the $D(1^3F_4)$ state indeed, it should be observed in both $D\pi$
and $D^*\pi$ channels. To clarify the puzzles in the $D^*_J(3000)$,
more observations are needed.

\begin{widetext}
\begin{center}
\begin{table}[ht]
\caption{The strong partial decay widths (MeV) and total width (MeV)
of $D^*(3000)$ as a candidate of $D(1^3F_4)$.} \label{Tab:D13F4}
\begin{tabular}{c| ccccccccc }\hline
\hline
 Mode \ \ &$D\pi$ &$D^*\pi$ &$D(2420)\pi$ &$D(2430)\pi$    &$D(2460)\pi$   &$D(2550)\pi$ &$D(2600)\pi$
 &$D(2750)\pi$ &$D(2760)\pi$  \\

 Width \ \ &14.9 & 11.0    & 2.3         &9.2     & 7.0   & 0.9
&$5.9\times10^{-2}$  & 4.6 &11.4  \\
\hline Mode \ \    &$D\eta$ &$D\eta'$ &$D^*\eta$
&$D^*\eta'$&$D(2420)\eta$ &$D(2430)\eta$ &$D(2460)\eta$
&$D_s(2460)K$
&$D_s^*K$\\
Width \ \ & 1.0 &$3.1\times10^{-2}$&0.5
&$8.9\times10^{-5}$&$5.6\times10^{-4}$ &$1.0\times10^{-3}$
&$8.1\times10^{-12}$&$3.6\times10^{-3}$
&0.8\\
\hline Mode \ \ &$D_sK$ &$D\omega$ &$D^*\omega$
   &$D\rho$&$D^*\rho$&$D_sK^*$&$D_s^*K^*$&Total &\\
Width \ \ &2.1 &$9.2\times10^{-6}$&4.5&1.6&14.5&$5.9\times10^{-7}$&$6.4\times10^{-4}$&86.4 &\\
\hline\hline
\end{tabular}
\end{table}
\end{center}
\end{widetext}


In the $D_s$-meson family, the predicted mass for the $1^3F_4$ state
is $3.22\sim3.30$ GeV~\cite{Di Pierro:2001uu,
Ebert:2009ua,Sun:2014wea}. The strong decay properties are shown in
Fig.~\ref{fig-3F4} as well. In the possible mass region, the
predicted width of the $D_s(1^3F_4)$ state is $\Gamma\simeq 200\pm
100$ MeV. The main decay channels are $D^*K^*$, $DK$, $D^*K$,
$D(2430)K$, and $D(2460)K$. If the mass of the $D_s(1^3F_4)$ state
is larger than the threshold of $D(2760)K$, the $D(2760)K$ channel
will become a dominant decay channel. The predicted partial width
ratio of $DK$ and $D^*K$ is
\begin{eqnarray}
\frac{\Gamma[DK]}{\Gamma[D^*K]}\simeq 1.2.
\end{eqnarray}
This state might be observed in both $DK$ and $D^*K$ channels.


In the $B$-meson family, the predicted mass for the $1^3F_4$ state
is $6.22\sim6.38$ GeV~\cite{Di Pierro:2001uu,
Ebert:2009ua,Sun:2014wea}. In this mass region, we study the strong
decay properties of the $B(1^3F_4)$ state, which are shown in
Fig.~\ref{fig-3F4}. From the figure, it is seen that the strong
decays of the $B(1^3F_4)$ state are dominated by the $B\pi$ and
$B^*\pi$, their partial width ratio is
\begin{eqnarray}
\frac{\Gamma[B\pi]}{\Gamma[B^*\pi]}\simeq 1.1,
\end{eqnarray}
which is less sensitive to the mass. Furthermore, the other decay
channels $B^*\rho$, $B(1^3D_3)\pi$, $B(1^2P_3)\pi$ and $B(1P_1)\pi$
also have obvious contributions to the strong decays. The decay
width is sensitive to the mass. In the possible mass region, the
predicted width of the $B(1^3F_4)$ state is $\Gamma\simeq 120\sim
480$ MeV. If this state has a mass of $\sim6.2$ GeV as predicted in
Ref.~\cite{Di Pierro:2001uu}, it might be observed in future
experiments. However, if the mass is close to $\sim6.4$ GeV as
predicted in Ref.~\cite{Ebert:2009ua}, the $B(1^3F_4)$ state might
be difficult to be found for its broad width.


In the $B_s$-meson family, the predicted mass for the $1^3F_4$ state
is $6.33\sim6.48$ GeV~\cite{Di Pierro:2001uu,
Ebert:2009ua,Sun:2014wea}. In this mass region, we have shown the
strong decay properties in Fig.~\ref{fig-3F4}. The decay width of
the $B_s(1^3F_4)$ is sensitive to the mass. The total decay width is
$\Gamma\simeq 60\sim 400$ MeV, which bares a large uncertainty. Its
strong decays are dominated by the $BK$, $B^*K$ and $B^*K^*$
channels. The predicted partial width ratio of the $BK$ and $B^*K$
is
\begin{eqnarray}
\frac{\Gamma[BK]}{\Gamma[B^*K]}\simeq 1.2,
\end{eqnarray}
which is insensitive to the mass. If this state has a smaller mass
of $\sim6.34$ GeV as predicted in Ref.~\cite{Di Pierro:2001uu}, it
should be a narrow state with a width of $\Gamma\simeq 80$ MeV. In
this case, the $B_s(1^3F_4)$ might be observed in both $BK$ and
$B^*K$ channels.

As a whole the decay widths of the $1^3F_4$ excitations in the
heavy-light mesons are sensitive to the mass. If these states have a
smaller mass as predicted in Ref.~\cite{Di Pierro:2001uu}, they
might have relatively narrow width $\Gamma\sim 100$ MeV. In this
case, these excitations have good discovery potentials in future
experiments. However, if these states have a larger mass as
predicted in Ref.~\cite{Ebert:2009ua}, they might be difficult to be
found in experiments for their broad widths. Finally, it should be
pointed out that the newly observed natural parity resonance
$D^*_J(3000)$ seems to be a good candidate of the $D(1^3F_4)$,
although its mass is about $100\sim 200$ MeV less than the model
predictions.

\section{Summary}\label{suma}

In the chiral quark model framework, we systematically study the
strong decays of the higher excited heavy-light mesons from the
first radially excited states up to the first $F$-wave states. We
summarize our major results as follows.


The first radially excited states $2^1S_0$ in the $D$-, $D_s$-, $B$-
and $B_s$-meson spectroscopy are narrow states, and their predicted
width might be less than 100 MeV. If the $D(2550)$ corresponds to
this excitation indeed, the other radially excited states
$D_s(2^1S_0)$, $B(2^1S_0)$ and $B_s(2^1S_0)$ are most likely to be
also observed in the $D^*K$, $B^*\pi$ and $B^*K$ final states,
respectively, for their narrower widths.


Configuration mixing might exist between the low-lying $J^P=1^-$
states $2^3S_1$ and $1^3D_1$. Both $D(2600)$ and $D_{s1}(2700)$
could be assigned as the low-mass mixed state $|(SD)_{1}\rangle_L$
via the $2^3S_1$-$1^3D_1$ mixing. As flavor partners of the
$D(2600)$ and $D_{s1}(2700)$, the low-mass mixed states in the $B$-
and $B_s$-meson families, $B(|(SD)_{1}\rangle_L)$ and
$B_s(|(SD)_{1}\rangle_L)$, should have a relatively narrow width.
They are most likely to be observed in future experiments. In the
high-mass mixed states, the widths of the charmed-strange state
$D_s(|(SD)_{1}'\rangle_H)$ and the bottom-strange state
$B_s(|(SD)_{1}'\rangle_H)$ are as comparable as the low-mass mixed
states, which might be observed in the $DK$ and $BK$ channels,
respectively. While, the bottom state $B(|(SD)_{1}'\rangle_H)$ is a
broad state, which might be difficult to be found in experiments. It
should be pointed out that the newly observed resonance
$D_{s1}(2860)$ by LHCb Collaboration might correspond to the
physical partners of the $D_{s1}(2700)$, i.e., the high-mass state
$D_s(|(SD)_{1}'\rangle_H)$.


In the low-lying $D$-wave states with $J^P=2^-$, $1^1D_2$ and
$1^3D_2$, there might be configuration mixing as well. The narrow
resonances $D(2750)$ and $D_{sJ_2}(2860)$ might be classified as the
high-mass mixed state $|1D_2'\rangle_H$ via the $1^1D_2$-$1^3D_2$
mixing. The other high-mass mixed states in the $B$- and $B_s$-meson
families, $B(|1D_2'\rangle_H)$ and $B_s(|1D_2'\rangle_H)$ have a
narrow width as well. These two narrow states might be observed in
the $B^*\pi$ and $B^*K$, respectively. However, the low-mass mixed
state $|1D_2\rangle_L$ should be a broad state. The typical total
decay widths for the mixed states $D(|1D_2\rangle_L)$ and
$B(|1D_2\rangle_L)$ in the $D$- and $B$-meson families are usually
larger than $300$ MeV, which might be too broad to be observed in
experiments. While the widths for the mixed states
$D_s(|1D_2\rangle_L)$ and $B_s(|1D_2\rangle_L)$ containing a strange
quark are $\sim 250$ MeV, which still have some possibilities to be
observed in future experiments.

For the low-lying $D$-wave excitations with $J^P=3^-$, $1^3D_3$, in
the $D$-, $D_s$-, $B$- and $B_s$-meson spectroscopy, we might have
observed three excitations $D(2760)$, $D_{sJ_1}(2860)$ and $B(5970)$
in recent experiments. The last unobserved one in the $B_s$-meson
family should be a narrow state with a width of $\Gamma\simeq
(25\sim 75)$ MeV, which is most likely to be found in the $BK$ and
$B^*K$ channels.

No evidence of the second $P$-wave excitations with $J^P=0^+$ is
found in experiments. In the $D$- and $B$-meson families, the
$D(2^3P_0)$ and $B(2^3P_0)$ excitations should be very broad states,
whose widths are larger than 400 MeV. It might be difficult to be
found in experiments. However, the widths of the resonances
$D_s(2^3P_0)$ and $B_s(2^3P_0)$ in the $D_s$- and $B_s$-meson
spectroscopy are relatively narrower, which might be observed in the
$DK$ and $BK$ channels, respectively.

The physical states of the second $P$-wave states with $J^P=1^+$
might be mixed states between the $2^1P_1$ and $2^3P_1$.  The strong
decay properties of both $D_{sJ}(3040)$ and $D_J(3000)$ can be
explained by assigning them as the low-mass mixed state
$|2P_1\rangle_L$ via the $2^1P_1$-$2^3P_1$ mixing with a mixing
angle in the range of $\phi_{2P}=-(20\sim26)^\circ$. The mixed state
$B_s(|2P_1\rangle_L)$ in the $B_s$-meson family is the narrowest
state in these low-mass mixed states, thus, it is most likely to be
observed in the $B^*K$ channel. On the contrary, the mixed state
$B(|2P_1\rangle_L)$ in the $B$-meson family might have the broadest
width in these low-mass mixed states, thus, its discovery potentials
might be small. The high-mass mixed states $D(|2P_1'\rangle_H)$,
$D_s(|2P_1'\rangle_H)$, $B(|2P_1'\rangle_H)$ and
$B_s(|2P_1'\rangle_H)$ are usually narrower than those of low-mass
states. These states dominantly decay into the first $P$-wave states
by emitting a light pseudoscalar meson. However, experimental
analysis of these decay channels is still absent, which might
explain why these broader mixed states $D_{sJ}(3040)$ and
$D_J(3000)$ have been first found in experiments. It is strongly
suggested to carry out experimental analysis of the final states
containing a low-lying $P$-wave state with $J^P=0^+,1^+$.

No evidence of the second $P$-wave excitations with $J^P=2^+$ in the
$D$-, $D_s$-, $B$- and $B_s$-meson spectroscopy is found in
experiments. Our calculations indicate that these $P$-wave
excitations $D(2^3P_2)$, $D_s(2^3P_2)$, $B(2^3P_2)$ and
$B_s(2^3P_2)$ have a relatively narrow width, their strong decays
are governed by the $D_1(2430)\pi$, $D_1(2430)K$, $B(1P_1)\pi$ and
$B(1P_1)K$, respectively. Thus, it is might be a good chance for us
to find the $D(2^3P_2)$ and $D_s(2^3P_2)$ excitations by analyzing
the data in the $D_1(2430)\pi$ and $D_1(2430)K$ final states,
respectively.

The $1^3F_2$ excitations in the heavy-light mesons are most likely
to be very broad resonances, whose widths are larger than 400 MeV.
Such broad states might be difficult to be observed in experiments,
which might explain why these states are still missing in the
heavy-light meson spectroscopy.

Considering configuration mixing in the first $F$-wave states with
$J^P=3^+$, we predict that the low-mass mixed states
$|1F_{3}\rangle_L$ in the heavy-light mesons are usually broad
states, they should be difficult to be found in experiments. While
the decay widths of the high-mass mixed states $|1F_{3}'\rangle_H$
are relatively narrower than those of the low-mass states. These
high-mass states $|1F_{3}'\rangle_H$ might be observed in future
experiments if they have a smaller mass as predicted in
Ref.~\cite{Di Pierro:2001uu}. The optimal observed channels for the
$D(|1F_{3}'\rangle_H)$, $D_s(|1F_{3}'\rangle_H)$,
$B(|1F_{3}'\rangle_H)$ and $B_s(|1F_{3}'\rangle_H)$ are $D^*\pi$,
$D^*K$, $B^*\pi$ and $B^*K$, respectively. However, these high-mass
states $|1F_{3}'\rangle_H$ might be very broad states if their mass
is as large as that predicted in Ref.~\cite{Ebert:2009ua}.

For the $1^3F_4$ excitations, their decay widths are sensitive to
the mass. If these states have a smaller mass as predicted in
Ref.~\cite{Di Pierro:2001uu}, they might have relatively narrow
width $\Gamma\sim 100$ MeV. In this case, these excitations have
good observation potentials in future experiments. The optimal
observed channels for the $D(1^3F_{4})$, $D_s(1^3F_{4})$,
$B(1^3F_{4})$ and $B_s(1^3F_{4})$ are $D\pi$/$D^*\pi$, $DK$/$D^*K$,
$B\pi$/$B^*\pi$ and $BK$/$B^*K$, respectively. However, if these
states have a larger mass as predicted in Ref.~\cite{Ebert:2009ua},
they might be difficult to be found in experiments for their broad
widths. Taking the newly observed natural parity resonance
$D^*_J(3000)$ as a candidate of the $D(1^3F_4)$, we find that the
theoretical predictions of the strong decay properties are
consistent with the observation. However, the predicted mass is
inconsistent with the observation, which is about $100\sim 200$ MeV
larger than the data. If the $D^*_J(3000)$ corresponds to the
$D(1^3F_4)$ state indeed, it should be observed in both $D\pi$ and
$D^*\pi$ channels. To clarify the puzzles in the $D^*_J(3000)$, more
observations are needed.

Finally, it should be pointed out that the chiral quark model still
has some limitations. For example, the simple harmonic oscillator
wave functions of the resonances have been adopted in the
calculations, which should be different from the realistic wave
functions more or less. Thus, the wave functions might bring some
uncertainties to our predictions. Furthermore, our model is a
nonrelativistic model, the relativistic effects could bring some
uncertainties to the decay widths as well. Thus, we might only give
a qualitative prediction of the strong decay properties for some
resonances.

\section*{ Acknowledgements }

This work is supported, in part, by the National Natural Science
Foundation of China (Grants No. 11075051 and No. 11375061), and the
Hunan Provincial Natural Science Foundation (13JJ1018).

\end{document}